\documentclass[english]{article}
\usepackage[T1]{fontenc}
\usepackage[applemac]{inputenc}
\usepackage{color}
\usepackage{babel}
\usepackage{graphicx}
\usepackage[left=2.4cm,top=2.2cm,right=2.4cm,bottom=2.2cm]{geometry}
\usepackage{epstopdf}
\usepackage{amsmath}
\usepackage{amssymb}
\usepackage{setspace}
\usepackage{esint}
\usepackage[unicode=true,pdfusetitle,
 bookmarks=true,bookmarksnumbered=false,bookmarksopen=false,
 breaklinks=false,pdfborder={0 0 1},backref=false,colorlinks=false]
 {hyperref}


\begin{document} 

\title{{\huge{}Probing Gravitational Cat States in Canonical Quantum Theory
vs Objective Collapse Theories}}

\author{Maaneli Derakhshani\thanks{Email: maanelid@yahoo.com and }}

\maketitle
\begin{center}
\emph{\large{}Institute for History and Foundations of Science \&
Department of Mathematics, Utrecht University, Utrecht, The Netherlands}
\par\end{center}{\large \par}

~
\begin{abstract}
Using as a testbed the recently proposed ``gravcat'' experimental
scheme in \cite{DerAnasHu2016}, we compare the properties of gravitational
cat states in three descriptions: (1) canonical quantum theory (CQT)
combined with the Newtonian limit of GR, (2) objective collapse theories
(OCTs) extended to the regime of semiclassical Newtonian gravity,
and (3) OCTs extended to incorporate quantized Newtonian gravity.
For the CQT approach, we follow the treatment by Hu and Anastopoulos
in \cite{Anastopoulos2015}. For the OCTs, we consider the GRW, CSL,
DP, and Karolyhazy theories, based on the semiclassical approaches
of Derakhshani \cite{Derakhshani2014} and Tilloy-Diósi \cite{Tilloy2016},
respectively, and we consider the most straightforward extension of
the aforementioned OCTs to the regime of quantized Newtonian gravity.
We show that the gravcat scheme can, in principle, experimentally
discriminate the quantum jumps in gravitational cat states predicted
by the CQT approach and the quantized-gravity OCTs (which we show
make effectively the same predictions as each other), from the predictions
of the semiclassical-gravitational OCTs. We also show that the GRW
and Karolyhazy versions of semiclassical gravity (based on Derakhshani's
approach) make distinctly different predictions from the CSL and DP
versions of semiclassical gravity (based either on Derakhshani's approach
or the Tilloy-Diósi approach). 
\end{abstract}

\section{Introduction}

Recent years have seen a flurry of papers analyzing the empirical
predictions of the Schrödinger-Newton equations \cite{W.Marshall2003,Adler2007,Diosi2007,Carlip2008,DomenicoGiulini2011,Meter2011,H.Yang2013,Giulini2013,Colin2014,M.Bahrami2015,Grossardt2015,Gan2016,Grossardt2016}
and canonical \footnote{By ``canonical'', we just mean the standard/textbook/ordinary version
of quantum theory. } quantum theory combined with the Newtonian limit of GR \cite{Anastopoulos2015,M.Bahrami2015,DerAnasHu2016},
for state-of-the-art AMO experiments designed to implement quantum
superpositions of mesoscopic masses \cite{M.Aspelmeyer2014,O.Romero-Isart2012,H.Pino2016}.
In parallel, a few papers \cite{Derakhshani2014,Nimmrichter2015,Tilloy2016}
in recent years have proposed consistent extensions of well-known
objective collapse theories (e.g., GRW, CSL, DP, etc.) to the regime
of semiclassical Newtonian gravity. What has yet to be done is an
assessment of the predictions of these semiclassical Newtonian gravity
versions of objective collapse theories, as well as objective collapse
theories extended to the regime of quantized Newtonian gravity, for
the proposed state-of-the-art AMO experiments. 

We contribute in this respect by working out the predictions of several
well-known objective collapse theories (OCTs), extended to semiclassical
Newtonian gravity and quantized Newtonian gravity (hereafter OCT-Newton
theories), for the gravitational cat state probe setup recently proposed
by Derakhshani, Anastopoulos, and Hu \cite{DerAnasHu2016}. Additionally,
we compare the predictions of these OCT-Newton theories to the predictions
of canonical quantum theory within the Newtonian approximation to
GR, the latter of which has been worked out by Anastopoulos \& Hu
in \cite{Anastopoulos2015} and used as the theoretical basis of the
grav-cat probe setup.

The paper is organized as follows. Section 2 reviews the grav-cat
scenario considered by Anastopoulos \& Hu as well as the grav-cat
probe setup of Derakhshani, Anastopoulos, and Hu. Section 3 reviews
and develops the semiclassical Newtonian gravity and quantized Newtonian
gravity extensions of the GRW, CSL, DP, Tilloy-Diósi, and Karolyhazy
objective collapse theories, and works out their predictions for the
grav-cat probe setup while pointing out where their predictions differ
from (or agree with) those of canonical quantum theory; secondarily,
these findings are used to swiftly assess related objective collapse
theories incorporating Newtonian gravity effects \cite{Adler2002,Adler2013,Kafri2014,Nimmrichter2015,Bera2015}.
Finally, section 4 summarizes and appraises our findings, and suggests
future research directions.

\section{GravCat states in canonical quantum theory}

Here we first review the general gravitational cat state scenario
examined by Anastopoulos \& Hu (AH) \cite{Anastopoulos2015}, then
the specific grav-cat setup proposed by Derakhshani, Anastopoulos,
and Hu (hereafter DAH) \cite{DerAnasHu2016}.

\subsection{General model}

Consider the canonical quantum theory (CQT) description of a stationary
point mass $M$ with initial (Gaussian) wavefunction 
\begin{equation}
\psi_{0}(\mathbf{x})=\frac{1}{\left(2\pi\sigma^{2}\right)^{3/4}}e^{-\frac{x^{2}}{4\sigma^{2}}}.
\end{equation}
In the canonical formalism, this wavefunction says that the position
$\mathbf{x}$ of the particle is a random variable with probability
density $|\psi_{0}(\mathbf{x})|^{2}$. By Newton's law, a probability
density for $\mathbf{x}$ entails a probability distribution for the
Newtonian force acting on a particle of mass $m$ at location $R$
as
\begin{equation}
\mathbf{F}=-\frac{GMm}{|\mathbf{R}-\mathbf{x}|^{3}}\left(\mathbf{R}-\mathbf{x}\right).
\end{equation}
For $|\mathbf{R}|\gg\sigma$, the quantum fluctuations of the Newtonian
force are negligible, and one recovers (effectively) the usual deterministic
Newtonian force. 

Suppose then that the wavefunction of the point mass $M$ is described
by a cat state, i.e., a superposition of two identical Gaussians,
each located at $\pm\frac{1}{2}\mathbf{L}$ and with zero mean momentum:
\begin{equation}
\psi_{cat}(\mathbf{x})=\frac{1}{\sqrt{2}}\frac{1}{\left(2\pi\sigma^{2}\right)^{3/4}}\left[e^{-\frac{(\mathbf{x}+\mathbf{L}/2)^{2}}{4\sigma^{2}}}+e^{-\frac{(\mathbf{x}-\mathbf{L}/2)^{2}}{4\sigma^{2}}}\right],
\end{equation}
Since the force is a function of $\mathbf{x}$, and $\mathbf{x}$
is described by a quantum operator, the Newtonian force must also
be an operator, and so should the corresponding gravitational potential.
Thus the cat state for the point mass generates a cat state for the
gravitational field. Moreover, when $\mathbf{L}$ is comparable to
$\mathbf{R}$, the fluctuations of Eq. (2) are non-negligible. 

Now suppose that a quantum particle of mass $M$, confined in a symmetric
potential depicted in Fig. 1, has two local minima at $\mathbf{r}=\pm\frac{1}{2}\mathbf{L}$,
labeled as $+$ and $-$. The general cat state is then given by 
\begin{equation}
|\psi>=c_{+}|+>+c_{-}|->,
\end{equation}
where $|+>$ and $|->$ are the state-vectors localized around the
corresponding minima. The system Hamiltonian is assumed to be $\hat{H}=\nu\hat{\sigma}_{1}$,
where $\nu$ is a small tunneling rate between the minima. 

\begin{figure}
\includegraphics[width=0.55\textwidth]{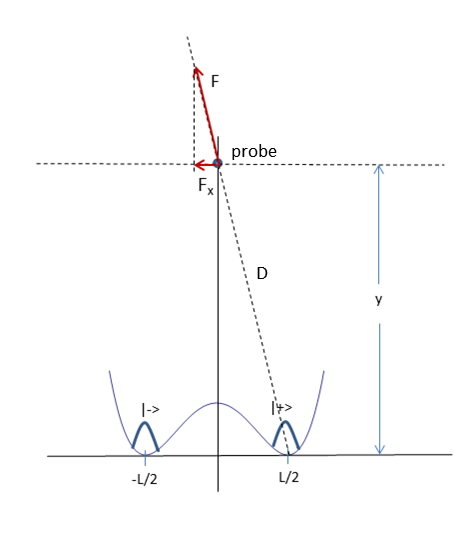}

\protect\caption{Force on a probe exerted by a massive particle in a gravitational
cat state.}
\end{figure}

To experimentally probe such a cat state, a classical probe and a
quantum probe were suggested and analyzed. It was determined that
the quantum probe is beyond foreseeable feasibility, so we will only
review the classical probe scheme. 

Consider a stationary test mass \emph{m} (the probe) located near
the confining potential as in Fig. 1. Assuming that the gravitational
force between the probe and the quantum particle causes the cat state
to collapse into one or the other of the minima, with probabilities
$|c_{+}|^{2}$ and $|c_{-}|^{2}$, the force \emph{F} in the horizontal
direction takes only two values, $f_{0}$ and $-f_{0}$, where 
\begin{equation}
f_{0}=\frac{GMmL}{2D^{3}},
\end{equation}
where $D=\sqrt{y^{2}+L^{2}/4}$ is the distance between the potential
minimum and the location of the probe; $y$ is shown in Fig. 1.

Then, it was shown in \cite{Anastopoulos2015} that for an initial
state $\left|+\right\rangle $, the quantum expectation value of $F$
and its two-time correlation function are given by 
\begin{equation}
\left\langle F(t)\right\rangle =-f_{0}e^{-\varGamma t},
\end{equation}
and 
\begin{equation}
\left\langle F(t')F(t)\right\rangle =f_{0}^{2}e^{-\varGamma|t'-t|}.
\end{equation}
The decay constant $\varGamma$ is defined as 
\begin{equation}
\Gamma=\frac{\nu^{2}\tau}{2},
\end{equation}
where $\tau$ is the probe's temporal resolution. 

From hereon, let us refer to the above as the gravitational cat state
(g-cat) scenario predicted by `CQT-Newton'.

Observe that in the above CQT-Newton treatment of the g-cat scenario,
the test mass is assumed to be a single particle. But what if the
test mass is, instead, a many-body system, such as a homogeneous sphere
composed of \emph{N} particles? How does CQT-Newton describe the gravitational
coupling of a many-body test mass with the classical probe, in the
limit that \emph{N} is large? AH \cite{AnastopoulosHu2014} have shown
that CQT-Newton, when understood as the Newtonian limit of the theory
of perturbatively quantized gravity, with \emph{N}-body Hamiltonian
(minus the renormalized mass term and assuming identical particles)
\begin{equation}
\hat{H}_{quant}=\hat{T}+\hat{U}_{int}=-\sum_{i=1}^{N}\frac{\hbar^{2}}{2m}\nabla_{i}^{2}-\sum_{i\neq j}\sum_{j}\frac{Gm^{2}}{|\hat{\mathbf{r}}_{i}-\hat{\mathbf{r}}_{j}|},
\end{equation}
has a mean-field (Hartree) approximation given by the single-body
Schrödinger-Newton (SN) equations
\begin{equation}
\nabla^{2}V_{int}=4\pi Gm|\chi(\mathbf{r},t)|^{2},
\end{equation}
and 
\begin{equation}
i\hbar\frac{\partial\chi(\mathbf{r},t)}{\partial t}=\hat{H}_{SN}\chi(\mathbf{r},t)=\left[-\frac{\hbar^{2}}{2m}\nabla^{2}-G\int d\mathbf{r}'\frac{m^{2}|\chi(\mathbf{r}',t)|^{2}}{|\mathbf{r}-\mathbf{r}'|}\right]\chi(\mathbf{r},t),
\end{equation}
if we assume that the \emph{N}-body wavefunction $\left|\Psi(t)\right\rangle =e^{-\left(\frac{i}{\hbar}\right)\hat{H}_{quant}t}\otimes_{i=1}^{N}\left|\chi\right\rangle =\otimes_{i=1}^{N}\left|\chi(t)\right\rangle $
for \emph{N} identical particles in the limit that $N\rightarrow\infty$.
It should be stressed, however, that despite formal similarities,
the physical interpretation of (10-11) is different from the physical
interpretation of the single-body SN equations treated as a fundamental
(Newtonian) description of the coupling of a single quantum particle
to gravity. Equations (10-11) describe the evolution of a collective
variable, $\chi(\mathbf{r},t)$ describing the large \emph{N} limit
of a Newtonian system of \emph{N} identical particles weakly interacting
via the perturbatively quantized gravitational potential $\hat{V}_{int}$. 

In any case, if we suppose, instead, that the test mass is a homogeneous
spherical distribution of \emph{N} identical particles described by
the Hamiltonian (9), then (10-11) should be a valid description of
the test mass. Moreover, if we place the N-body test mass near the
double well potential illustrated in Fig. 1, we simply add $U_{well}$
to the right side of (11), and the initial state $\left|\chi\right\rangle $
takes the cat state form (4). Then, (10-11) take the form 
\begin{equation}
\nabla^{2}V_{int}=4\pi Gm\left[|c_{+}|^{2}|\chi_{+}(\mathbf{r},t)|^{2}+|c_{-}|^{2}|\chi_{-}(\mathbf{r},t)|^{2}\right],
\end{equation}
and 
\begin{equation}
i\hbar\frac{\partial\chi(\mathbf{r},t)}{\partial t}=\left[-\frac{\hbar^{2}}{2m}\nabla^{2}+U_{well}-G\int d\mathbf{r}'\frac{m^{2}|c_{+}|^{2}|\chi_{+}(\mathbf{r},t)|^{2}}{|\mathbf{r}-\mathbf{r}'|}-G\int d\mathbf{r}'\frac{m^{2}|c_{-}|^{2}|\chi_{-}(\mathbf{r},t)|^{2}}{|\mathbf{r}-\mathbf{r}'|}\right]\chi(\mathbf{r},t),
\end{equation}
where $\left|\chi(t)\right\rangle =e^{-\frac{i}{\hbar}\hat{H}_{SN}t}\left|\chi\right\rangle $,
$\chi_{+}(\mathbf{r},t)=\left\langle \mathbf{r}\right.\left|\chi_{+}(t)\right\rangle $,
and $\chi_{-}(\mathbf{r},t)=\left\langle \mathbf{r}\right.\left|\chi_{-}(t)\right\rangle $
(there will also be an interaction term between $+$ and $-$, which
we will neglect for simplicity). If we maintain the assumption of
a classical probe, then (12-13) say that the probe will feel a classical
Newtonian gravitational force from classical mass densities localized
around each minimum of the potential with proportions $|c_{+}|^{2}$
and $|c_{-}|^{2}$ of the total test mass \emph{m}, respectively.
Moreover, if we suppose that $|c_{+}|^{2}=|c_{-}|^{2}=\frac{1}{2}$,
then the net force on the probe will clearly be zero, in contradiction
to what CQT-Newton predicts for the case of a single particle test
mass. One might attempt to resolve this inconsistency by simply postulating
that the classical gravitational coupling of the probe to the cat
state causes $\left|\chi\right\rangle $ to collapse into either $\left|\chi_{+}\right\rangle $
or $\left|\chi_{-}\right\rangle $, with probabilities $|c_{+}|^{2}=|c_{-}|^{2}=\frac{1}{2}$.
However, it has been pointed out by numerous authors \cite{Adler2007,J.vanWezel2008,Derakhshani2014,AnastopoulosHu2014}
that (12-13), whether interpreted as a fundamental theory or a mean-field
theory, has no consistent Born-rule probability interpretation. So
it would seem that the large \emph{N} limit of CQT-Newton makes a
prediction for the g-cat scenario that differs significantly from
the single particle case of CQT-Newton. We might conclude from this
that the mean-field approximation leading to (12-13) is simply not
valid for cat states, or that some other assumption in the mean-field
approximation doesn't hold for the g-cat setup. In this regard, we
can make two observations: the mean-field approximation leading to
(12-13) is designed to be valid only when quantum fluctuations of
the matter degrees of freedom are small \cite{B.L.Hu2008,Ford2005}
\footnote{However, in a forthcoming paper, we will show that even if we incorporate
the back-reaction of the quantum fluctuations of $\chi(\mathbf{r},t)$
on $V_{int}$ via the Newtonian limit of Hu and Verdaguer's stochastic
gravity theory \cite{B.L.Hu2008}, this does not yield a prediction
for the g-cat setup that's in better agreement with the exact CQT-Newton
description.}, which is clearly not the case for the cat state of the g-cat setup
described in Fig. 1; and (2) the ansatz that particle correlations
are negligible so that $\left|\Psi(t)\right\rangle =\otimes_{i=1}^{N}\left|\chi(t)\right\rangle $
will be a poor approximation if the test mass is (say) a solid spherical
body of uniform density composed of \emph{N} identical particles (which
will indeed be the case for the experimental protocols we shall consider
in subsection 2.2). 

Thus, for the description of a solid homogeneous spherical test mass
composed of \emph{N} identical particles and placed in the cat state
illustrated in Fig. 1, we should stick to the exact quantum description
corresponding to the Hamiltonian (9), but with the addition of an
interaction potential $\hat{U}_{int}^{ele}(\hat{\mathbf{r}}_{i}-\hat{\mathbf{r}}_{j})$
reflecting the non-gravitational (e.g., electrostatic) binding forces
between the particles. Then we should rewrite the Hamiltonian in terms
of the center of mass (CM) coordinate of the test mass (which means
we can drop $\hat{U}_{int}$ and $\hat{U}_{int}^{ele}$, since they
only contribute to the relative mass Hamiltonian), add $U_{well}$,
and add a gravitational potential $-\frac{GMm_{tot}}{\hat{D}}$ describing
the g-coupling of the CM of the test mass with the CM of the classical
probe. Putting it all together, we have the CM Schrödinger equation
\begin{equation}
i\hbar\frac{\partial\psi(\mathbf{r}_{cm},t)}{\partial t}=\left[-\frac{\hbar^{2}}{2m_{tot}}\nabla_{cm}^{2}-\frac{GMm_{tot}}{\hat{D}}+U_{well}\right]\psi(\mathbf{r}_{cm},t),
\end{equation}
where $m_{tot}=Nm$, $\hat{D}=\sqrt{y_{cm}^{2}+|\mathbf{\hat{x}}_{cm}|^{2}}$,
$\mathbf{\hat{x}}_{cm}$ has eigenvalues $-\mathbf{L}/2$ and $+\mathbf{L}/2$,
and $\mathbf{y}_{cm}$ is the fixed (c-number) y-displacement of the
CM of the test mass from the CM of the probe. With this description
in hand, we can take (4) as the initial state of the CM wavefunction
in (14) and straightforwardly apply AH's single particle analysis,
thereby reaching their same general conclusions.

Let us now revisit the CQT-Newton assumption that the gravitational
force interaction with the classical probe `causes' the quantum particle's
cat state to collapse into a definite position eigenstate with Born-rule
probabilities. It might ask why this should be so, as opposed to the
classical gravitational field of the probe acting as a mere external
field that weakly perturbs the Hamiltonian of the quantum particle.
Moreover, if the gravitational force measurement by the probe does
act as a projective measurement, how can experimentally fashioned
cat states of any mass remain stable at all (which they evidently
can \cite{Arndt2014}), given the presence of other (and much more
massive) gravitating bodies such as the Earth? (The usual assumption
in the application of the canonical quantum measurement postulates
to experiments is that the physical coupling of the system to the
pointer variable is the strongest coupling in the experiment, which
is clearly not the case here.)

These questions run right into the well-known quantum measurement
problem that afflicts CQT, insofar as CQT is intrinsically vague about
exactly what kinds of physical interactions in nature constitute projective
measurements \footnote{To quote John Bell, ``It would seem that the theory {[}quantum mechanics{]}
is exclusively concerned about `results of measurement', and has nothing
to say about anything else. What exactly qualifies some physical systems
to play the role of `measurer'? Was the wavefunction of the world
waiting to jump for thousands of millions of years until a single-celled
living creature appeared? Or did it have to wait a little longer,
for some better qualified system ... with a Ph.D.? If the theory is
to apply to anything but highly idealized laboratory operations, are
we not obliged to admit that more or less `measurement-like' processes
are going on more or less all the time, more or less everywhere. Do
we not have jumping then all the time?'' \cite{Bell1990}.}, exactly when projective measurements occur in or outside experiments,
exactly where they occur in the (so-called) von-Neumann chain of an
experiment, and exactly what dynamical laws govern the state-vector
reduction process (as opposed to unitary evolution) \cite{Bell1990,AdlerMP2003,Schlosshauer2004,Bassi2000}.
Nevertheless, if we view canonical quantum theory as a convenient
operationalist formalism (i.e., a formalism about agents and how they
can extract information from the microscopic physical world in experiments),
then we might say \footnote{This perspective was suggested by Charis Anastopoulos (private communication).}
the following: CQT requires, in order for a projective measurement
to occur, that there exists a well-defined macroscopic pointer variable
that agents can use to extract information from microscopic systems
they experimentally couple to the pointer variable. Insofar as the
classical probe is designed to be such a macroscopic pointer, it is
reasonable to predict that the gravitational force measurement will
indeed play the role of a projective measurement. By contrast, for
other massive gravitating bodies in nature (e.g., Earth), it is difficult
to see what could play the role of an appropriate pointer variable,
so we have no justification (within operationalist CQT) for expecting
that gravitational force interactions between the quantum particle
and (say) the Earth will collapse the latter's cat state.

Of course, this raises the question what `information extraction'
and `macroscopic' mean, exactly. While we concur that these notions
needs further elaboration, for the purposes of this paper, we will
not pursue the issue. Rather, we will take it as a working assumption
that the gravitational force interaction with the classical probe
(and only the classical probe) plays the role of a projective measurement,
in accordance with the usual measurement postulates of CQT.

\subsection{Experimental setup}

Within the framework of CQT-Newton, DAH proposed an experimental scheme
to actually measure the gravitational force between a classical probe
\footnote{We shall forego analysis of a quantum probe since DAH found that the
use of a quantum probe makes it far too difficult to measure the gravitational
force.} and a massive quantum particle in a cat state. For preparing the
g-cat state, they primarily considered Romero-Isart et al.'s \cite{O.Romero-Isart2012}
proposed experimental protocol involving the use of a superconducting
lead (Pb) microsphere (the quantum particle) of $M\sim10^{14}amu$
and radius $R=2\mu m$, which is first trapped (via Meissner effect)
in a harmonic potential created by a magnetic quadrupole, then parametrically
coupled to a qubit circuit to put the microsphere into a spatial superposition
of $\mathbf{L}\sim1pm$. For the role of the classical probe, it was
decided that the most promising experimental implementation is Reinhardt
et al.'s \cite{Sankey2015} trampoline resonator made from $Si_{3}N_{4}$, with effective mass $m=4.0 ng$, width $100\mu m$, and projected force
sensitivity of $\sim14zN$ at cryogenic temperatures ($14mK$). 

While the resonator is a square-like membrane rather than a point
particle, Eq. (5) can be used for an order of magnitude estimate of
the force. For a resonator of mass $m=4.0ng$, a microsphere of mass
$M=0.38ng$, $\mathbf{L}=1pm$ and $D=3\mu m$ (or $1\mu m$ larger
than the radius of the Pb microsphere), we obtain
\begin{equation}
f_{0}=\frac{GmML}{2D^{3}}\sim2\times10^{-30}N,
\end{equation}
which is around ten orders of magnitude beyond the reach of the projected
force sensitivity range of the resonator. To examine the optimal means
by which to enhance the resonator\textendash microsphere gravitational
interaction, DAH write $D=R+a$, where $R$ is the radius of the microsphere
and $a$ is a fixed distance between the surface of the sphere and
the resonator (we will consider $a\sim1\mu m$). Then
\begin{equation}
f_{0}\simeq(2)\frac{G\rho_{lead}mL}{\left(1+a/R\right)^{3}},
\end{equation}
where $\rho_{lead}=M/\left(\frac{4}{3}\pi R^{3}\right)=11.36\frac{g}{cm^{3}}$
is the density of the microsphere. From this we can see that the most
important parameter to increase the force is the cat state size $\mathbf{L}$,
followed by the radius $R$, then the density $\rho_{lead}$ of the
microsphere (consideration of Casimir forces \cite{Mohideen1998}
puts a practical lower bound on $a\geq1\mu m$). With these considerations,
DAH showed that if we can increase the size of the cat $\mathbf{L}$
by one order of magnitude, use a Tantalum microsphere of density $\rho_{tantalum}=16.7\frac{g}{cm^{3}}$,
and assume that $R=5\mu m$ is feasible for a Tantalum microsphere,
we obtain
\begin{equation}
f_{0}=0.6\times10^{-28}N,
\end{equation}
or still about eight orders of magnitude from the peak sensitivity
of the resonator. 

However, DAH suggested other possibilities for upping the sphere-resonator
force, such as increasing the mass of the resonator (though the importance
of the gravitational self-energy of the probe would then have to be
assessed). Another is to use a different protocol for preparing a
microsphere in a cat state, since further increases in the $R$ (hence
$M$) of the microsphere in Romero-Isart et al.'s protocol are limited
by decoherence from trap fluctuations \cite{O.Romero-Isart2012,Romero-IsartUnpublishedmanuscript};
in particular, Pino et al.'s \cite{H.Pino2016} recently proposed
protocol involving free expansion in a magnetic skatepark potential
(see Figure 3 therein), which makes possible a microsphere mass of
$M\sim10^{13}amu$ ($R\gtrsim1\mu m$) with $\mathbf{L}\sim500nm$
or more (since trap fluctuations are significantly lessened by the
free expansion). With this value for $\mathbf{L}$ the above force
estimates would increase by five orders of magnitude or more, i.e.,
$\sim10^{-25}N$ for the initial assessment, and $\sim10^{-23}N$
for the second assessment. Thus DAH concluded that ``the quantum
effects of a matter source manifested through its gravitational field
interactions could become measurable in the next (or next-next) generation
of experiments'' \cite{DerAnasHu2016}.

We should note that an experimental implementation of the double well
potential was not discussed by DAH, which limits our analysis to only
an estimation of the probe-sphere gravitational force interaction
when the sphere is initially prepared in the cat state (4). \footnote{The state-of-the-art method for experimentally fashioning double well
potentials appears to be the use of optical tweezers, which can produce
double-well minima spacings as small as $600nm$ \cite{A.M.Kaufman2014}.
It seems implausible that optical tweezers (or any other method) could
produce well-defined minima spacing of $1pm$. But it seems not far
off to produce minima spacings of $500nm$, which is relevant for
the Pino et al. protocol. }

Given the well-known conceptual ambiguities associated with the nature
of measurement in CQT (the measurement problem), and the speculative
nature of extrapolating the standard quantum measurement postulates
to the gravitational field, it is natural to ask how alternative quantum
theories that unambiguously solve the measurement problem might change
the predicted outcomes of the above g-cat setup. Since, to date, the
only alternative quantum theories that have been consistently extended
to the regime of semiclassical Newtonian gravity are objective collapse
theories \cite{G.C.Ghirardi1986,Karolyhazy1986,Diosi1987,Diosi1989,GhirardiPearle1990,Adler2002,A.Bassi2013,Adler2013,Derakhshani2014,Nimmrichter2015,Bera2015,Tilloy2016},
we will consider them specifically.

\section{GravCat states in objective collapse theories}

Here we analyze and compare the predictions of the most well-known
and well-defined objective collapse theories that have been extended
to semiclassical Newtonian gravity, and compare their predictions
to those of CQT-Newton, for the g-cat setup considered in the previous
section. Then we do the same for objective collapse theories extended
to incorporate quantized Newtonian gravity.

\subsection{Collapse theories with semiclassical gravity}

\subsubsection{GRW}

Among objective collapse theories the mathematically simplest one
is the GRW theory \cite{G.C.Ghirardi1986,A.Bassi2013}, based as it
is on the Poisson process. Likewise, among existing objective collapse
theories that have been extended to the regime of semiclassical Newtonian
gravity, the mathematically and conceptually simplest one appears
to be the GRW-Newton (hereafter GRWmN) theory of Derakhshani \cite{Derakhshani2014}.
Let us briefly review this approach. 

For a single-body system, we postulate the existence of an ontic matter
density field in space-time,

\begin{equation}
m(\mathbf{x},t)=m|\psi(\mathbf{x},t)|^{2},
\end{equation}
which is used as a source in the Newton-Poisson equation,

\begin{equation}
\nabla^{2}V(\mathbf{x},t)=4\pi Gm(\mathbf{x},t),
\end{equation}
where

\begin{equation}
V(\mathbf{x},t)=-G\int\frac{m(\mathbf{x}',t)}{|\mathbf{x}-\mathbf{x}'|}d^{3}\mathbf{x}'.
\end{equation}
This gravitational `self-potential' couples back to the wavefunction
via the Schrödinger-Newton (SN) equation, 
\begin{equation}
i\hbar\partial_{t}\psi(x,t)=-\frac{\hbar^{2}}{2m}\nabla^{2}\psi(\mathbf{x},t)-Gm\int d^{3}\mathbf{x}'\frac{m(\mathbf{x}',t)}{|\mathbf{x}-\mathbf{x}'|}\psi(\mathbf{x},t),
\end{equation}
but now the wavefunction undergoes discrete and instantaneous intermittent
collapses according to the GRW collapse law. That is, the collapse
time $T$ occurs randomly with constant rate per system of $N\lambda_{GRW}=\lambda_{GRW}=10^{-16}\frac{1}{s}$,
where the post-collapse wavefunction $\psi_{T+}=lim_{t\searrow T}\psi_{t}$
is obtained from the pre-collapse wavefunction $\psi_{T-}=lim_{t\nearrow T}\psi_{t}$
through multiplication by a Gaussian function,

\begin{equation}
\psi_{T+}(\mathbf{x})=\frac{1}{C}g(\mathbf{x}-\mathbf{X})^{1/2}\psi_{T-}(\mathbf{x}),
\end{equation}
where

\begin{equation}
g(\mathbf{x})=\frac{1}{(2\pi\sigma^{2})^{3/2}}e^{-\frac{x^{2}}{2\sigma^{2}}}
\end{equation}
is the 3-D Gaussian function of width $\sigma_{GRW}=10^{-7}m$, and

\begin{equation}
C=C(\mathbf{X})=\left(\int d^{3}xg(\mathbf{x}-\mathbf{X})|\psi_{T-}(\mathbf{x})|^{2}\right)^{1/2}
\end{equation}
is the normalization factor. The collapse center $\mathbf{X}$ is
chosen randomly with probability density $\rho(\mathbf{x})=C(\mathbf{x})^{2},$
and the space-time locations of the collapses are given by the ordered
pair $\left(\mathbf{X}_{k,}T_{k}\right).$ Between collapses, the
wavefunction evolves by (18-21). 

The generalization to an \emph{N}-body system is as follows. We have
\emph{N} matter density fields in 3-space,

\begin{equation}
m(\mathbf{x},t)=\sum_{i=1}^{N}\int d\mathbf{y}{}_{1}...d\mathbf{y}{}_{N}|\psi(\mathbf{y}{}_{1},...,\mathbf{y}{}_{N},t)|^{2}m_{i}\delta^{(3)}(\mathbf{x}-\mathbf{y}{}_{i}),
\end{equation}
which act as the mass density source in the Newton-Poisson equation,
\begin{equation}
\nabla^{2}V(\mathbf{x},t)=4\pi G\sum_{i=1}^{N}\int d\mathbf{y}{}_{1}...d\mathbf{y}{}_{N}|\psi(\mathbf{y}{}_{1},...,\mathbf{y}{}_{N},t)|^{2}m_{i}\delta^{(3)}(\mathbf{x}-\mathbf{y}{}_{i}).
\end{equation}
The solution of (26) couples back to the \emph{N}-body wavefunction
via

\begin{equation}
i\hbar\partial_{t}\psi(\mathbf{x}_{1}...\mathbf{x}_{N},t)=-\underset{i=1}{\overset{N}{\sum}}\frac{\hbar^{2}}{2m_{i}}\nabla_{i}^{2}\psi(\mathbf{x}_{1}...\mathbf{x}_{N},t)-G\sum_{i,j=1}^{N}\int\frac{m_{i}m_{j}(\mathbf{x}_{j}',t)}{|\mathbf{x}_{i}-\mathbf{x}_{j}'|}d\mathbf{x}'_{1}...d\mathbf{x}'_{N},
\end{equation}
and the solution of (27) undergoes collapse according to

\begin{equation}
\psi_{T+}(\mathbf{x}_{1},...,\mathbf{x}_{N})=\frac{1}{C}g(\mathbf{x}_{i}-\mathbf{X})^{1/2}\psi_{T-}(\mathbf{x}_{1},...,\mathbf{x}_{N}),
\end{equation}
with probability density

\begin{equation}
\rho(\mathbf{X})=C(\mathbf{X})^{2}=\int d\mathbf{x}'_{1}...d\mathbf{x}'_{N}g(\mathbf{x}'_{i}-\mathbf{X})|\psi_{T-}(\mathbf{x}'_{1},...,\mathbf{x}'_{N})|^{2},
\end{equation}
where $i$ is chosen randomly from $1,...,N.$

The equations of \emph{N}-body GRWmN say the following: the wavefunction
propagates on configuration space $\mathbb{R}^{3N}$, evolves by the
many-body SN equations, (26-27), and undergoes the collapse process
in (28-29); this wavefunction drives the dynamical evolution of \emph{N}
matter density fields in 3-space via (25) so that when the wavefunction
collapses, it localizes the matter density fields around randomly
chosen (non-overlapping) points in 3-space, each of width $10^{-7}$
meters, with rate $N\lambda_{GRW}$, and with probability density
given by (29). As before, each of these matter density fields acts
as a source for a classical Newtonian gravitational potential in 3-space
that couples back to the \emph{N}-body wavefunction via (26-27), which
in turn alters the evolution of the matter density fields via (25)
again. As shown in \cite{Derakhshani2014}, this dynamics suppresses
macroscopic gravitational cat states and has a consistent single-particle
probabilistic interpretation, in contrast to the SN equations alone
\cite{Diosi1984,Adler2007,AnastopoulosHu2014,Diosi2016}. 

Insofar as GRWmN is based on the SN equations, the nonlinearity of
the theory makes it possible, in principle, to do superluminal signaling
using (for example) spin-1/2 particles passing through a Stern-Gerlach
apparatus. However, the signaling effect is so tiny that it is well-beyond
present experimental capabilities to detect \footnote{Bahrami et al. \cite{BahramiBassi2014} point out that for state-of-the-art
experiments, which can achieve quantum interference with $m\sim10^{4}amu$
and Stern-Gerlach detectors with a spatial separation of $d\sim1\mu m$,
the minimum distance on which a Stern-Gerlach experiment would need
to be carried out to do superluminal signaling is 1 light-year.} \cite{BahramiBassi2014}. Thus, while it might be regarded by some
as a philosophically undesirable feature of the theory (and certainly
one in inherent conflict with special relativity), it doesn't seem
to entail empirical inadequacies at the present time \footnote{If one is worried about superluminal signals creating causal paradoxes
in different Lorentz frames \cite{Gisin1989,Diosi2016}, this can
be eliminated through the introduction of a preferred foliation of
spacetime \cite{Tumulk2006} such as the foliations already used in
relativistic flat-space extensions of ordinary GRW and CSL \cite{Bedingham2011,BedinghamTumulka2012}. }. 

We now apply GRWmN to the g-cat setup. To do this, we model the center
of mass of the microsphere cat state with the single-body GRWmN equations,
and we consider the case of the classical force probe (see again Fig.
1). In this case, for the initial cat state 
\begin{equation}
\psi_{cat}(\mathbf{x})=\frac{1}{\sqrt{2}}\frac{1}{\left(2\pi\sigma^{2}\right)^{3/4}}\left[e^{-\frac{(\mathbf{x}+\mathbf{L}/2)^{2}}{4\sigma^{2}}}+e^{-\frac{(\mathbf{x}-\mathbf{L}/2)^{2}}{4\sigma^{2}}}\right],
\end{equation}
which we assume is confined to a symmetric double-well potential with
the two local minima located at $\mathbf{r}=\pm\mathbf{L}/2$, measurement
probabilities $|c_{+}|^{2}=|c_{-}|^{2}=\frac{1}{2}$, and such that
the overlap between the summands in (30) is negligible (which is reasonable
if we make the barrier potential sufficiently large), we then have
the cat state matter density
\begin{equation}
M_{cat}(\mathbf{x})=M|\psi_{cat}(\mathbf{x})|^{2}=\frac{1}{2}\frac{M}{\left(2\pi\sigma^{2}\right)^{3/2}}\left[e^{-\frac{(\mathbf{x}+\mathbf{L}/2)^{2}}{2\sigma^{2}}}+e^{-\frac{(\mathbf{x}-\mathbf{L}/2)^{2}}{2\sigma^{2}}}\right]=\frac{M_{+}(\mathbf{x})}{2}+\frac{M_{-}(\mathbf{x})}{2}.
\end{equation}
Here $M$ is taken to be the mass of the microsphere cat state, and
the terms $\frac{1}{2}M_{+}(\mathbf{x})$ and $\frac{1}{2}M_{-}(\mathbf{x})$
describe lumps of halved center-of-mass microsphere matter densities
localized around the position eigenvalues of the left and right minima
of the potential, respectively. Thus, for a classical probe of mass
$m$ located at a distance $D=\sqrt{y^{2}+L^{2}/4}$ from the two
minima, the horizontal force that the probe feels from the two matter
densities is 
\begin{equation}
\mathbf{f}=\frac{GMmL}{4D^{3}}\mathbf{\hat{L}}-\frac{GMmL}{4D^{3}}\mathbf{\hat{L}}=0.
\end{equation}
That is, at every instant in time, the probe feels no net gravitational
force from the cat state. 

Note that this prediction will only hold when either (i) each collapse
event negligibly changes the cat state wavefunction, or (ii) the number
of particles composing the microsphere does not imply an appreciable
collapse rate, and the gravitational coupling of the probe to the
cat state does not drive up the collapse rate of the cat state.

Are these conditions satisfied by GRWmN applied to the g-cat setup? 

In the case of (i), recall that in the Romero-Isart et al. protocol
the microsphere cat size $\mathbf{L}=10^{-12}m$. By comparison, $\sigma_{GRW}=10^{-7}m$.
Thus, GRW collapses will leave the microsphere cat state unchanged,
if the cat state is prepared with the Romero-Isart et al. protocol.
In the Pino et al. protocol, by contrast, the microsphere cat size
$\mathbf{L}\gtrsim5\times10^{-7}m$; so in this case, GRW collapses
will appreciably localize the microsphere cat state. 

In the case of (ii), recall that, in the GRW formalism, the rate of
collapse for a many-body system scales (under the simplest assumption)
as $N\lambda_{GRW}$. Now, what determines that a given system is
`comprised of \emph{N} particles' is that the particles are interacting
strongly enough that their interaction Hamiltonian is non-negligible
and implies a non-separable many-body wavefunction $\psi(\mathbf{x}_{1},...,\mathbf{x}_{N})$.
Certainly the microspheres used in the Romero-Isart et al. and Pino
et al. protocols satsify this condition. So for a microsphere mass
of $M\sim10^{14}amu$, we have $N\lambda_{GRW}\sim10^{14}\cdot10^{-16}\frac{1}{s}=10^{-2}\frac{1}{s}$,
or one collapse event every 100 seconds. Since 100 seconds is a timescale
well beyond any feasible timescale of the g-cat experiment, whether
using the Romero-Isart et al. protocol or the Pino et al. protocol
for preparing the cat state, we can ignore this collapse rate for
microspheres of said mass. Furthermore, it is clear that the semiclassical
gravitational coupling (as described by the SN equations) between
the microsphere matter density and the probe's matter density \emph{will
not} drive up the microsphere's collapse rate, simply because the
semiclassical gravitational coupling does not lead to an entanglement
between the many-body wavefunction of the microsphere cat state and
the many-body wavefunction of the probe; that is, the semiclassical
gravitational coupling implies that the probe's wavefunction and associated
matter density do not evolve into a superposition of orthogonal pointer
states that are correlated with the position states of the microsphere
cat state (i.e., the two minima of the double well potential) \footnote{This can be seen by noting that since the initial matter densities
imply no net gravitational deflection of the probe in the x-direction,
time-evolving the sphere-probe system with the Hamiltonian (27) will
not evolve the probe's wavefunction into effectively orthogonal pointer
states.}. Instead, all that the semiclassical gravitational coupling does
is introduce a slight phase shift in the many-body wavefunction of
the microsphere. Moreover, the same will be true of semiclassical
gravitational coupling between the microsphere and any other matter
density present in the experiment (e.g., the Earth, the sun, etc.). 

Additionally, it is easy to see that the gravitational self-energy
of the (un-collapsed) cat state matter density is negligible in both
protocols since the interaction energy between the two lumps of matter
in the two minima is given by $U_{self}=-\frac{GM^{2}}{4L}$, where
$U_{self}|_{L=1pm}\sim10^{-35}J$ and $U_{self}|_{L=500nm}\sim10^{-41}J$.
By contrast, a potential barrier between the minima of just $1eV\approx10^{-19}J$.
In other words, the cat state matter density will remain stably confined
in the double well potential.

As noted in section 2, the Pino et al. protocol \footnote{The protocol involves seven steps \cite{H.Pino2016}: (1) \textbf{Cooling}.
Cooling the center-of-mass (CM) motion of the superconducting microsphere
in a harmonic trap of frequency $\omega_{1}$ for a time $t_{1}$
to a definite phonon occupation number; (2) \textbf{Boost}. Evolving
the CM in an inverted harmonic potential of frequency $\omega_{2}$
for a time $t_{2}$ in order to boost the sphere's kinetic energy;
(3) \textbf{Free}. Free evolution for a time $t_{3}$ to delocalize
the CM over long distances; (4) \textbf{Split}. Continuous-time measurement
of the position-squared for a time $t_{4}$ to implement diffraction
through a double-slit `barrier' (where the measurement outcome determines
the slit separation, and the strength of the measurement determines
the width of the slits) and prepare a quantum spatial superposition
state; (5) \textbf{Rotation}. Short evolution for a time $t_{5}$ in
a harmonic trap of frequency $\omega_{5}$ to give opposite momenta
to the wavepackets in the superposition; (6) \textbf{Inflation}. Evolution
in an inverted quadratic potential of frequency $\omega_{6}$ for
a time $t_{6}$ to exponentially generate interference fringes; and
(7) \textbf{Measurement}. A measurement of the position-squared for
a time $t_{7}$ to unveil the interference pattern. } affords us the possibility of even larger microsphere masses (hence
larger collapse rates) than does the Romero-Isart et al. protocol.
Suppose then that we assume the Pino et al. protocol for preparing
our microsphere cat state and that it allows us a sphere mass as large
as $\sim10^{18}amu$. (We will forego a discussion of the practical
details of how to experimentally implement the double-well potential,
but the general idea is that each minimum of the potential would be
located around one of the slits in the double-slit barrier, so that
when the microsphere matter density emerges from the slits as two
distinct lumps of half-mass matter densities, each lump will be trapped
in one of the minima, as in Fig. 1.) Then the intrinsic collapse rate
of the microsphere will be $10^{-2}\frac{1}{s}$, or 1 collapse event
every millisecond. By comparison, in the Pino et al. protocol, the
total time between preparing and detecting a coherent microsphere
cat state of $10^{13}amu$ with $\mathbf{L}=0.5\mu m$ is the sum
over the time intervals for steps 2-4 in the protocol, or $\sim500ms$.
For a microsphere of $\sim10^{18}amu$, it will presumably take at
least this long to form a coherent cat state in the protocol. In this
case, the coherence time of the microsphere will in fact \emph{exceed}
the inverse collapse rate of the microsphere. More precisely, the
coherent microsphere would undergo dozens of collapse events before
being put into a cat state via diffraction through the two-slit `barrier',
and dozens of collapse events thereafter. Note that, in the latter
case, each collapse event would correspond to multiplying the cat
state wavefunction, which essentially takes the form (30) with the
peaks of the Gaussians separated by $0.5\mu m$, by the Gaussian function
(22), giving 

\begin{equation}
\psi_{T+}(\mathbf{x})=\frac{1}{C}g(\mathbf{x}-\mathbf{X})^{1/2}\psi_{cat}(\mathbf{x}),
\end{equation}
which is a wavepacket of width $\sigma=0.1\mu m$ with collapse center
probability

\begin{equation}
C=C(\mathbf{X})=\left(\int d^{3}\mathbf{x}g(\mathbf{x}-\mathbf{X})|\psi_{cat}(\mathbf{x})|^{2}\right)^{1/2},
\end{equation}
where $\mathbf{X}=\{+\frac{\mathbf{L}}{2},-\frac{\mathbf{L}}{2}\}$. 

Here we should elaborate on what happens to the microsphere wavepacket
and matter density immediately after a GRW collapse event. As we've
noted, in between collapse events, the wavepacket resumes its evolution
via the \emph{N}-body SN equations, where the gravitational interaction
potential clearly causes the \emph{N}-body wavepacket to self-gravitate.
Will this self-gravitation be sufficiently strong to keep the collapsed
wavepacket from dispersing again? This was studied numerically by
Giulini \& Grossardt \cite{Giulini2013}, who found that for a homogeneous
sphere of mass\emph{ }$M$ and initial radius $R=0.5\mu m$, the density
of the corresponding spherically symmetric center-of-mass (CM) wavepacket
begins to undergo ``gravitational collapse'' at a mass as low as
$\sim5\times10^{9}amu$, and reaches a minimum radius in a time of
$\sim20,000s$. For $R=1\mu m$, the critical mass for gravitational
collapse to set in is $\sim8\times10^{9}amu$. And for the maximum
simulated mass of $\sim10^{11}amu$ ($R=0.5\mu m$), the shortest
gravitational collapse duration was observed to be $\sim2,200s$.
Giulini \& Grossardt were unable to simulate larger masses due to
numerical limitations, so it is not possible to say by how much more
the gravitational collapse time would be reduced for $\sim10^{18}amu$
and $R=1\mu m$. But it seems implausible that it would be reduced
to a timescale of $\sim500ms$. In any case, these results make it
clear that the SN self-interaction is strong enough to override the
quantum mechanical wavepacket dispersion, for a microsphere with $R=1\mu m$
or less. 

So when a GRW collapse occurs and localizes the microsphere CM wavepacket
to the width $\sigma_{GRW}=0.1\mu m$, we can be sure that the SN
self-interaction will prevent the wavepacket from delocalizing again.
Moreover, any subsequent GRW collapse event will make no change to
the CM wavepacket width. Thus, when a GRW collapse event occurs before
the microsphere is split into two lumps via the double-slit (step
4 in the protocol), we can predict that when the microsphere does
finally interact with the double-slit (where the slit width $w=10.61nm$
and slit separation $d=0.5\mu m$), it will either get reflected or
pass through one or the other of the slits, emerging on the other
side as a full-mass lump trapped within one or the other of the double-well
minima. In other words, the GRW collapse coupled with the SN self-interaction
will actually make it unfeasible to put the microsphere into a cat
state with the Pino et al. protocol. Consequently, the gravitational
force continuously detected by the probe will come from only one of
the minima and remain so, as the SN self-interaction and GRW collapses
will inhibit tunneling between the minima. 

By contrast, CQT-Newton predicts that even for $M\sim10^{18}amu$
and $R=1\mu m$, it should be possible to form the microsphere cat
state with the Pino et al. protocol, and the gravitational force interaction
with the probe will instantaneously collapse the cat state CM wavefunction
into one of the minima, with probabilities $|c_{+}|=|c_{-}|=\frac{1}{2}$.
Additionally, continuous monitoring of the gravitational field by
the probe will still find the force undergoing quantum jumps between
the minima, as described by Eqs. (20-22).

We conclude then that it is necessary to adopt the Pino et al. protocol,
in order to have a chance of increasing the microsphere mass to a
level where (i) the probe-microsphere force becomes detectable, and
(ii) the probe-microsphere net force predicted by GRWmN becomes non-zero
and has correlation functions that distinctly differ from those of
CQT-Newton \footnote{We should note that our conclusions change nontrivially if we consider
GRWfN in place of GRWmN \cite{Derakhshani2014}. In GRWfN, each ``flash''
(i.e., space-time collapse center $\left\{ X_{k},T_{k}\right\} $)
is accompanied by the sudden appearance of a point mass at the space-time
location of the flash, and in between collapse events the wavefunction
evolves by the usual linear Schrödinger equation. Thus, when no flashes
are present in space-time, no matter density is present in space-time,
and no gravitational interactions can be present. So then, for the
g-cat setup, if the collapse rate is too low for the timescale of
the experiment, the probe will detect no gravitational force from
the microsphere cat state because there will be no mass density in
space-time associated with the cat state. On the other hand, if the
collapse rate is sufficiently high that dozens of collapse events
can occur on the timescale of the experiment, then we can predict
the following for the Pino et al. protocol: the GRW collapses of the
microsphere wavepacket that occur before it reaches the double-slit
barrier will be immediately followed by delocalization of the wavepacket
due to Schrödinger evolution. If a collapse occurs just before the
wavepacket interacts with the double-slit barrier, then the wavepacket
will be sufficiently narrow that it can only pass through one of the
slits (whichever one it is localized near), thus emerging into only
one of the minima (say the $+$ one) on the other side. From thereon,
the probe will detect an instantaneous force from the $+$ minimum
only when a GRW collapse event happens, followed by no force until
another GRW collapse happens. Since, in between GRW collapses, the
wavepacket evolves by the usual Schrödinger equation, it is possible
in these intermittent times that the wavepacket tunnels from the $+$
minimum to the $-$ minimum. So when the next GRW collapse event happens,
the probe might feel an instantaneous force coming from the - minimum,
followed by no force again until the next GRW collapse (which might
still come from the $-$ minimum or change back to the $+$ minimum).
On the other hand, if the wavepacket is still delocalized as it is
interacting with the double-slit barrier, then the cat state wavefunction
will form on the other side and remain that way until a GRW collapse
occurs. From hereon, the gravitational force changes in time as already
described. In either case, the predictions differ noticeably from
both CQT-Newton and GRWmN. }

\subsubsection{CSL}

As is well-known, the CSL theory is a quantum field theoretic generalization
of the GRW theory \cite{GhirardiPearle1990,BassiDeckertFerialdi2010,A.Bassi2013}
based (in its simplest formulation) on a continuous-time Markov process,
namely, the Wiener process. For our first approach, we consider a
straightforward semiclassical gravitational generalization of CSL
analogous to GRWmN (hereafter CSLmN). In particular, we adopt from
the mass-proportional version of non-relativistic CSL the stochastic
Schrödinger equation for the $N$-particle sector of Fock space:
\begin{equation}
\begin{split}\frac{\partial}{\partial t}\left|\psi_{t}\right\rangle  & =\left[-\frac{i}{\hbar}\hat{H}+\frac{\sqrt{\gamma}}{m}\int d^{3}\mathbf{x}\left(\hat{m}(\mathbf{x})-<\hat{m}(\mathbf{x})>_{t}\right)dW_{t}(\mathbf{x})\right.\\
 & \left.-\frac{\gamma}{2m^{2}}\int\int d^{3}\mathbf{x}d^{3}\mathbf{y}g(\mathbf{x}-\mathbf{y})\left(\hat{m}(\mathbf{x})-<\hat{m}(\mathbf{x})>_{t}\right)\left(\hat{m}(\mathbf{y})-<\hat{m}(\mathbf{y})>_{t}\right)\right]\left|\psi_{t}\right\rangle ,
\end{split}
\end{equation}
where 
\begin{equation}
\hat{H}=\hat{H}_{0}+\hat{U}_{g-int}
\end{equation}
is the system Hamiltonian, a sum of the usual kinetic energy operator
and gravitational interaction energy operator. The term
\begin{equation}
\hat{m}(\mathbf{y})=\frac{1}{m_{nuc}}\int d^{3}\mathbf{y}g(\mathbf{x}-\mathbf{y})\sum_{s}m_{s}a_{s}^{\dagger}(\mathbf{y})a_{s}(\mathbf{y})
\end{equation}
is the smeared spatial mass density operator, defined in terms of
the number density operator $a_{s}^{\dagger}(\mathbf{y})a_{s}(\mathbf{y})$,
where the sum is over particle species $s$ of mass $m_{s}$. The
parameter $m_{nuc}$ is the nucleon mass, and $g(\mathbf{x}-\mathbf{y})$
is a spatial correlation function chosen equal to the 3-D Gaussian
\begin{equation}
g(\mathbf{x})=\frac{1}{(2\pi r_{c}^{2})^{3/2}}e^{-\frac{x^{2}}{2r_{c}^{2}}}.
\end{equation}

Now, we define the semiclassical gravitational potential via the Poisson
equation
\begin{equation}
\nabla^{2}V_{g-int}(\mathbf{x},t)=4\pi G<\hat{m}(\mathbf{x})>_{t}
\end{equation}
with 
\begin{equation}
<\hat{m}(\mathbf{x})>_{t}=<\psi_{t}|\hat{m}(\mathbf{x})|\psi_{t}>,
\end{equation}
where it is readily confirmed that (40) is equivalent to the \emph{N}-body
mass density field (25), making the $V_{g-int}(\mathbf{x},t)$ of
(39) equivalent to the $V(\mathbf{x},t)$ of (26). Then the semiclassical
gravitational interaction energy (which includes self-interaction)
is given by
\begin{equation}
\hat{U}_{g-int}=\int d^{3}\mathbf{x}V_{g-int}(\mathbf{x},t)\hat{m}(\mathbf{x})=-G\int d^{3}\mathbf{x}\int d^{3}\mathbf{y}\frac{\hat{m}(\mathbf{x})<\hat{m}(\mathbf{y})>_{t}}{|\mathbf{x}-\mathbf{y}|},
\end{equation}

The noise term $dW_{t}(\mathbf{x})/dt=w(\mathbf{x},t)$ satisfies
\begin{equation}
\mathrm{E}\left[w(\mathbf{x},t)\right]=0,
\end{equation}
and
\begin{equation}
\mathrm{E}\left[w(\mathbf{x},t_{1})w(\mathbf{y},t_{2})\right]=g(\mathbf{x}-\mathbf{y})\delta(t_{2}-t_{1}),
\end{equation}
where $\mathrm{E}\left[...\right]$ is the stochastic average. Thus
the noise field is independent of the gravitational self-interaction,
affording us a consistent probabilistic interpretation of the spontaneous
localization of of the wavefunction. We also note that the fundamental
parameter $\gamma=10^{-36}m^{3}s^{-1}$ \cite{A.Bassi2013}, and the
collapse rate $\lambda_{CSL}$ is related to $\gamma$ by 
\begin{equation}
\lambda_{CSL}=\frac{\gamma}{8\pi^{3/2}r_{c}^{3}}\approx10^{-17}s^{-1},
\end{equation}
choosing the correlation length $r_{c}=10^{-7}m$. In CSL, spatial
superpositions separated by more than $r_{c}$ are localized with
effective rate \cite{A.Bassi2013}
\begin{equation}
\varGamma=\lambda_{CSL}N^{2}k,
\end{equation}
where $N$ is the number of particles within a distance $r_{c}$ and
$k$ is the number of such clusters of particles. (This dependence
on $N^{2}$ turns out to depend on the identity of the particles \cite{A.Bassi2013}.)
Accordingly, in a molecule or microsphere where the inter-particle
distances are much smaller than $r_{c}$, the collapse only affects
the center of mass motion, implying $\varGamma_{cm}=\lambda_{CSL}N^{2}=\lambda_{CSL}\left(\frac{m}{m_{nuc}}\right)$,
where $m$ is the total mass and $m_{nuc}$ is the mass of a nucleon. 

Applying CSLmN to the g-cat setup, it follows again that the probe-microsphere
gravitational interaction, defined here by
\begin{equation}
\hat{U}_{g}^{probe-sphere}=-\frac{G}{2}\int d^{3}\mathbf{x}\int d^{3}\mathbf{y}\frac{m_{probe}(\mathbf{x})<\hat{m}_{sphere}(\mathbf{y})>_{t}}{|\mathbf{x}-\mathbf{y}|},
\end{equation}
does not drive up the collapse rate of the microsphere, only the number
of nucleons composing the microsphere do so. So for $m_{sphere}\sim10^{14}amu$
we have $N\sim10^{14}$ nucleons, giving an effective localization
rate on the center of mass motion of $\varGamma_{cm}=\lambda N^{2}=10^{11}\frac{1}{s}$.
As with GRW/GRWmN, this collapse rate will have no effect on the cat
state formed by the Romero-Isart et al. protocol; but for the Pino
et al. protocol, this collapse rate means that for CSL/CSLmN, it will
not be possible, in practice, to experimentally prepare a coherent
and stable microsphere cat state in the form of (4) or (30) using
Pino et al.'s free expansion protocol (in contrast to GRW/GRWmN and
CQT/CQT-Newton). More precisely, the microsphere cat state formed
by the Pino et al. protocol will very quickly be suppressed by the
first collapse event that happens, and the SN self-interaction of
the collapsed microsphere matter density will ensure that the microsphere
stays localized in the minimum in which it got localized. So, were
force probe sensitive enough to measure the gravitational force from
$m_{sphere}\sim10^{18}amu$, it would only detect a force from a full-mass
microsphere matter density that's continuously localized around one
of the two minima in the g-cat setup, with zero probability of tunneling
between the minima in between collapse events (because of the appreciable
SN self-interaction). Clearly this prediction for the gravitational
force interaction would be experimentally indistinguishable from what
GRWmN predicts (when a collapse event happens in GRWmN before the
microsphere passes through the double-slit and forms into a cat state),
but sharply differs from what CQT-Newton predicts. Finally, we note
that these findings about CSLmN also apply to the dissipative generalization
of CSL (hence dissipative CSLmN) proposed by Smirne \& Bassi \cite{SmirneBassi2015},
insofar as the dissipative terms they incorporate don't alter the
CSL collapse rate.

\subsubsection{DP}

The DP theory \cite{Diosi1987,Diosi1989,Penrose1996,Penrose1998,M.BahramiBassi2014}
is structurally equivalent to mass-proportional CSL in that the equation
of motion for $\left|\psi_{t}\right\rangle $ is given by Eq. (35),
just with the replacements $\sqrt{\gamma}/m\rightarrow1$ and $\gamma/2m^{2}\rightarrow1/2$.
The key physical difference of DP is in the choice of spatial correlation
function:
\begin{equation}
g(\mathbf{x})=\frac{G}{\hbar}\frac{1}{|\mathbf{x}|}.
\end{equation}
In the density matrix formulation of DP, one then has 
\begin{equation}
\frac{\partial}{\partial t}\hat{\rho}(t)=-\frac{i}{\hbar}\left[\hat{H},\hat{\rho}(t)\right]+\frac{G}{\hbar}\int\int\frac{d^{3}\mathbf{x}d^{3}\mathbf{y}}{|\mathbf{x}-\mathbf{y}|}\left[\hat{m}(\mathbf{x})\hat{\rho}(t)\hat{m}(\mathbf{y})-\frac{1}{2}\left\{ \hat{m}(\mathbf{x})\hat{m}(\mathbf{y}),\hat{\rho}(t)\right\} \right].
\end{equation}

The different choice of spatial correlation function in DP entails
divergences in (48) which Diósi proposed to remedy with a length-scale
cut-off $R_{0}$ \cite{Diosi1987,Diosi1989,M.BahramiBassi2014}; however,
even with the cut-off, the energy of a system of particles increases
monotonically, and one also has the problem of overheating \cite{A.Bassi2013,M.BahramiBassi2014}.
As shown by Bahrami et al. \cite{M.BahramiBassi2014}, this overheating
problem can be dealt with by adding dissipative terms to (48) (including
the cut-off), but the overheating is still appreciable unless one
also requires an upper limit on the temperature of the noise field
in DP; in particular, Bahrami et al. deduce that the dissipative terms
lead to an asymptotic value of the noise field energy corresponding
to a temperature 

\begin{equation}
T=\frac{\hbar^{2}}{8k_{B}}\frac{1}{m_{r}R_{0}^{2}}=\frac{10^{-19}}{m_{r}R_{0}^{2}}.
\end{equation}
Here $m_{r}$ corresponds to the center of mass consistent with the
asymptotic value of $T$. Diósi proposed the choice of cut-off $R_{0}=10^{-15}m$
(corresponding to the Compton wavelength of a nucleon), which for
Bahrami et al.'s (physically reasonable) choice of $T=1K$ gives $m_{r}\sim10^{11}amu$.
For $m\ll m_{r}$, Bahrami et al. \cite{M.BahramiBassi2014} find
that the dissipation mechanism becomes too strong, leading them to
conclude that the dissipative DP theory is valid only as an effective
theory for masses comparable to or larger than $m_{r}$. 

The regularized version of Eq. (48) can be readily solved in the single-particle
case for a two-state spatial superposition. Consider the density matrix
$\left\langle \mathbf{x}\right|\hat{\rho}(t)\left|\mathbf{x}'\right\rangle $,
where $\mathbf{x}$ and $\mathbf{x}'$ are the two distinct locations
in the superposition. For short timescales, one can neglect the pure
Schrödinger contribution in (48) and solve to obtain 
\begin{equation}
\rho(\mathbf{x},\mathbf{x}',t)=exp\left(-\frac{t}{\tau(\mathbf{x},\mathbf{x}')}\right)\rho(\mathbf{x},\mathbf{x}',0),
\end{equation}
where for $|\mathbf{x}-\mathbf{x}'|\gg R_{0}$, one finds that the
characteristic damping time
\begin{equation}
\tau(\mathbf{x},\mathbf{x}')=\frac{\hbar}{\triangle E}\approx\frac{1}{\Lambda_{DP}}=\frac{\sqrt{\pi}\hbar R_{0}}{Gm^{2}},
\end{equation}
where $\triangle E$ is the Newtonian gravitational self-energy of
the massive particle at superposed locations $\mathbf{x}$ and $\mathbf{x}'$,
and $\Lambda_{DP}$ is the gravitational decoherence rate (the rate
at which the spatial superposition decays). 

Thus, for Romero-Isart et al.'s protocol involving a Pb microsphere
of $m\sim10^{14}amu\sim10^{-13}kg$, placed into a center of mass
spatial superposition of $|\mathbf{x}_{cm}-\mathbf{x}_{cm}'|\sim10^{-12}m$
via parametric coupling to a qubit, and using Diósi's choice of $R_{0}=10^{-15}m,$
Eq. (51) gives 
\begin{equation}
\tau_{sphere}(\mathbf{x},\mathbf{x}')\sim10^{-13}s.
\end{equation}
So the dissipative DP theory of Bahrami et al. predicts suppression
of the superposition on a timescale several orders of magnitude smaller
than any feasible timescale for the experimental setups we've considered.
A straightforward conclusion to draw, then, is that the dissipative
DP theory predicts that, for a microsphere with $m\sim10^{14}amu$,
a cat state of CM position states formed by either parametric coupling
to a qubit (Romero-Isart et al. protocol) or by free expansion in
a magnetic skatepark potential (Pino et al. protocol) will be rapidly
suppressed to a width of $10^{-15}m,$ on the timescale given by (52). 

If we assume a semiclassical gravitational extension of the dissipative
DP theory via the SN approach, i.e. (dissipative) DPmN, then for the
g-cat setup, continuous monitoring of the gravitational field of the
microsphere (confined to a double-well potential) by a classical probe
will result only in a force from the microsphere located in one of
the two minima of the potential for all times, as one would expect
classically. Insofar as this prediction holds for the g-cat setup
involving a microsphere cat state prepared by the Romero-Isart et
al. protocol, this prediction of dissipative DPmN is indistinguishable
from CSLmN; and for the g-cat setup involving a microsphere cat state
prepared by the Pino et al. protocol, the prediction is the same as
both CSLmN and GRWmN.

It is interesting to compare this result with Pino et al.'s analysis
of the DP theory using their experimental proposal \cite{H.Pino2016}.
They consider the original DP theory with only the cutoff $R_{0}$
and calculate a gravitational decoherence timescale of $\sim10^{-2}s$
for a microsphere of radius $R_{0}=1\mu m$ and center of mass $m\sim10^{13}amu$.
Clearly, then, the dissipative DP/DPmN theory is even more easily
falsifiable than the original DP theory.

\subsubsection{Tilloy-Diósi}

Because of its nonlinear dynamics, CSLmN implies superluminal signaling
just as GRWmN. Again, however, the effect is too small to measure
with state-of-the-art technology but might nevertheless be considered
philosophically unpalatable. Motivated by this philosophical dissatisfaction,
Tilloy \& Diósi \cite{Tilloy2016} developed a semiclassical Newtonian
gravitational extension of CSL that eliminates the nonlinearity that
implies superluminal signaling. In contrast to CSLmN (and GRWmN),
their theory implies only inter-particle gravitational potentials
and no single-particle self-interaction that depends on the wavefunction.
In addition, their theory contains as special cases the CSL theory
and the Diósi-Penrose theory, each amended with the inclusion of inter-particle
gravitational potentials. It is therefore worthwhile to also assess
the predictions of the Tilloy-Diósi approach for the experimental
setups considered here.

In order to circumvent the superluminal signaling entailed by using
$<\hat{m}(\mathbf{x})>_{t}$ as the source of the Newtonian gravitational
potential, Tilloy and Diósi (TD) propose to use a mass density source
defined from a fictitious model of hidden (and possibly entangled)
detectors of spatial resolution $\sigma$ that continuously monitor
the mass density operator $\hat{m}(\mathbf{x})$, the latter defined
as in Eq. (37). That is, they use the continuous equivalent of a von-Neumann
measurement result, i.e., the ``signal'' defined by 
\begin{equation}
m_{t}(\mathbf{x})=<\hat{m}(\mathbf{x})>_{t}+\delta m_{t}(\mathbf{x}),
\end{equation}
where $\delta m_{t}(\mathbf{x})$ is a spatially correlated white
noise field defined by
\begin{equation}
\mathrm{E}\left[\delta m_{t_{2}}(\mathbf{x})\delta m_{t_{1}}(\mathbf{y})\right]=\gamma_{\mathbf{xy}}^{-1}\delta(t_{2}-t_{1}),
\end{equation}
where $\gamma_{\mathbf{xy}}$ is a semi-positive definite kernel encoding
correlations between the fictitious detectors at positions $\mathbf{x}$
and $\mathbf{y}$. In order to implement these assumptions into a
continuous stochastic localization theory, TD suppose that, for an
\emph{N}-body system, the dynamics of the \emph{N}-body density matrix
$\hat{\rho}$ is defined by the stochastic master equation (SME)
\begin{equation}
\frac{d\hat{\rho}}{dt}=-i\left[\hat{H},\hat{\rho}\right]-\int d^{3}\mathbf{x}d^{3}\mathbf{y}\frac{\gamma_{\mathbf{xy}}}{8}\left[\hat{m}_{\sigma}(\mathbf{x}),\left[\hat{m}_{\sigma}(\mathbf{y}),\hat{\rho}\right]\right]+\int d^{3}\mathbf{x}d^{3}\mathbf{y}\frac{\gamma_{\mathbf{xy}}}{2}\mathcal{H}\left[\hat{m}_{\sigma}(\mathbf{x})\right]\hat{\rho}\delta m(\mathbf{y}),
\end{equation}
where $\mathcal{H}\left[\hat{m}_{\sigma}(\mathbf{x})\right]\left(\hat{\rho}\right)=\left\{ \hat{m}_{\sigma}(\mathbf{x})-\left\langle \hat{m}_{\sigma}(\mathbf{x})\right\rangle _{t},\hat{\rho}_{t}\right\} $
and we set $\hbar=1$. The deterministic term involving the double-commutator
describes the decoherence induced by the coupling with the fictitious
detectors and diagonalizes the density matrix in the position basis
for large-mass objects. The stochastic term induces localization of
the density matrix into one of its diagonal components, as a result
of the conditioning on the signal.

Now, in order to implement the back-action of quantized matter on
the classical gravitational field, TD define the Poisson equation
for the mass density signal, $m_{t}(\mathbf{x})$: 
\begin{equation}
\nabla^{2}V_{g-int}(\mathbf{x})=4\pi Gm_{t}(\mathbf{x}).
\end{equation}
Accordingly, the stochastic semiclassical gravitational self-interaction
energy associated with the signal is given by 
\begin{equation}
\hat{U}_{g-int}=\int d^{3}\mathbf{x}V_{g-int}(\mathbf{x})\hat{m}_{(\sigma)}(\mathbf{x})=\int d^{3}\mathbf{x}\hat{m}(\mathbf{x})V_{g-int(\sigma)}(\mathbf{x})
\end{equation}
where the subscript $(\sigma)$ denotes an optional convolution with
the smearing function, $g_{\sigma}$. The feedback from Eq. (57) to
$\hat{\rho}$ is introduced by having the self-interaction energy
act an infinitesimal amount of time after the free-evolution given
by Eq. (55), i.e.,
\begin{equation}
\hat{\rho}+d\hat{\rho}=e^{-i\hat{U}_{g-s}dt}\left(\hat{\rho}+d\hat{\rho}^{free}\right)e^{i\hat{U}_{g-s}dt}.
\end{equation}
Then, expanding the exponential in (58) up to second order, TD obtain
the SME

\begin{equation}
\begin{split}\frac{d\hat{\rho}}{dt} & =-i\left[\hat{H}+\hat{U}_{g,\sigma}+\int d^{3}x\delta m(\mathbf{x})\hat{V}_{(\sigma)},\hat{\rho}\right]\\
 & -\int d^{3}\mathbf{x}d^{3}\mathbf{y}\left(\frac{\gamma_{\mathbf{xy}}}{8}\left[\hat{m}_{\sigma}(\mathbf{x}),\left[\hat{m}_{\sigma}(\mathbf{y}),\hat{\rho}\right]\right]+\frac{\gamma_{\mathbf{xy}}^{-1}}{2}\left[\hat{V}_{(\sigma)}(\mathbf{x}),\left[\hat{V}_{(\sigma)}(\mathbf{y}),\hat{\rho}\right]\right]\right)\\
 & +\int d^{3}\mathbf{x}d^{3}\mathbf{y}\frac{\gamma_{\mathbf{xy}}}{2}\mathcal{H}\left[\hat{m}_{\sigma}(\mathbf{x})\right]\hat{\rho}\delta m(\mathbf{y}),
\end{split}
\end{equation}
where 
\begin{equation}
\hat{U}_{g,\sigma}=\frac{1}{2}\int d^{3}\mathbf{x}\hat{m}_{\sigma}(\mathbf{x})\hat{V}_{(\sigma)}(\mathbf{x})=-\frac{G}{2}\int d^{3}\mathbf{x}d^{3}\mathbf{y}\frac{\hat{m}_{\sigma}(\mathbf{x})\hat{m}_{(\sigma)}(\mathbf{y})}{|\mathbf{x}-\mathbf{y}|}
\end{equation}
is the Newtonian gravitational pair-potential up to $\sigma$-smearing
of the mass density around the point-like constituents of the localization
events. Here we can see how TD's proposal avoids nonlinearity: the
self-interaction of each signal only shifts the \emph{N}-body system
Hamiltonian by finite amounts and have no dynamical consequences.
We also see that the gravitational back-action induces an additional
local decoherence term that depends on $\hat{V}_{(\sigma)}(\mathbf{x})$.
Finally, we note that the stochastic term that drives the localization
of the density matrix remains the same as in the free-evolution case. 

Let us now examine the CSL case of TD's theory. This corresponds to
the spatial correlator

\begin{equation}
\gamma_{\mathbf{xy}}=\gamma\delta(\mathbf{x}-\mathbf{y}),
\end{equation}
along with the values $\sigma=10^{-7}m$ and $\gamma=10^{-24}m^{3}s^{-1}$
(though other values for these parameters are possible). Additionally,
since in CSL it is possible to define (57) in terms of the sharp mass
density without getting infinities, this is done by TD too (which
amounts to dropping the $\sigma$ subscripts in $\hat{U}_{g-int}$).
The resulting SME takes the form 

\begin{equation}
\begin{split}\frac{d\hat{\rho}}{dt} & =-i\left[\hat{H}+\hat{U}_{g,\sigma}+\int d^{3}\mathbf{x}\delta m(\mathbf{x})\hat{V},\hat{\rho}\right]\\
 & -\int d^{3}\mathbf{x}\left(\frac{\gamma}{8}\left[\hat{m}_{\sigma}(\mathbf{x}),\left[\hat{m}_{\sigma}(\mathbf{x}),\hat{\rho}\right]\right]+\frac{1}{2\gamma}\left[\hat{V}(\mathbf{x}),\left[\hat{V}(\mathbf{x}),\hat{\rho}\right]\right]\right)\\
 & +\int d^{3}\mathbf{x}\frac{\gamma}{2}\mathcal{H}\left[\hat{m}_{\sigma}(\mathbf{x})\right]\hat{\rho}\delta m(\mathbf{y}).
\end{split}
\end{equation}
We can now apply TD's version of CSL (TD-CSL) to the g-cat setup.
The term $\hat{U}_{g,\sigma}$ in (62) describes the gravitational
interaction energy between the probe and the microsphere; treating
the probe as a classical mass density, $m_{probe}(\mathbf{x})$, we
have 
\begin{equation}
\hat{U}_{g,\sigma}^{probe-sphere}=-\frac{G}{2}\int d^{3}\mathbf{x}d^{3}\mathbf{y}\frac{m_{probe}(\mathbf{x})\hat{m}_{sphere}(\mathbf{y})}{|\mathbf{x}-\mathbf{y}|},
\end{equation}
which will just contribute a phase shift to $\hat{\rho}$ as an external
field, but otherwise will not change the CSL collapse rate $\varGamma$.
As in CSLmN, the collapse rate will only depend on the number of particles
composing the microsphere and so will also yield $\varGamma_{cm}=\lambda N^{2}=10^{11}\frac{1}{s}$
for $m_{sphere}\sim10^{14}amu$. We can also calculate the effect
of the decoherence term due to gravitational back-action, in the special
case of a single particle of mass $m$ and density matrix $\hat{\rho}(\mathbf{x}_{1},\mathbf{x}_{2})$,
namely,

\begin{equation}
\begin{split}\hat{D}_{g}\left[\hat{\rho}\right] & =-\int d^{3}\mathbf{x}\frac{1}{2\gamma}\left[\hat{V}(\mathbf{r}),\left[\hat{V}(\mathbf{r}),\hat{\rho}\right]\right]\\
 & =-\frac{G^{2}m^{2}}{8\gamma}\int d^{3}\mathbf{r}\left(\frac{1}{|\mathbf{r}-\mathbf{x}_{1}|}-\frac{1}{|\mathbf{r}-\mathbf{x}_{2}|}\right)^{2}\rho(\mathbf{x}_{1},\mathbf{x}_{2})\\
 & =-\frac{\pi G^{2}m^{2}}{2\gamma}|\mathbf{x}_{1}-\mathbf{x}_{2}|\rho(\mathbf{x}_{1},\mathbf{x}_{2}).
\end{split}
\end{equation}
This expression tells us that the back-action decoherence term damps
the phases of the density matrix in proportion to the distance $|\mathbf{x}_{1}-\mathbf{x}_{2}|$
separating the position $\mathbf{x}_{1}$ and $\mathbf{x}_{2}$ corresponding
to the two possible locations of the signal. If we take $|\mathbf{x}_{1}-\mathbf{x}_{2}|\sim10^{-6}m$,
which is applicable to the Pino et al. protocol, we obtain 
\begin{equation}
\hat{D}_{g}\left[\hat{\rho}\right]\sim-10^{-29}\rho(\mathbf{x}_{1},\mathbf{x}_{2}),
\end{equation}
indicating extremely slow phase damping. For $|\mathbf{x}_{1}-\mathbf{x}_{2}|\sim10^{-12}m$,
which is applicable to the Romero-Isart et al. protocol, it is obvious
the phase damping rate is even slower (but we can ignore this case
since, as we know, the CSL collapses will not destroy a cat state
with this small a distance separating $\mathbf{x}_{1}$ and $\mathbf{x}_{2}$).
So we can conclude that TD-CSL also predicts that it should not be
possible to experimentally prepare stable microspheres with (centers
of) mass $\sim10^{14}amu$ in coherent spatial superpositions using
Pino et al.'s protocol. Moreover, like CSLmN, TD-CSL predicts that
for the g-cat setup, a classical probe will detect a virtually constant
force from a microsphere (prepared with the Pino et al. protocol)
that's virtually continuously localized in one of the two minima of
the double-well potential. We say ``virtually'' because, unlike
CSLmN, no SN self-interaction is present, so there is still a small
probability of each component of the cat state tunneling between the
minima between collapse events, but it seems unlikely to be observable
on realistic timescales of the g-cat experiment. 

For the DP case of TD's theory (TD-DP), the correlator is less trivial:

\begin{equation}
\gamma_{\mathbf{xy}}=\kappa G\frac{1}{|\mathbf{x}-\mathbf{y}|},
\end{equation}
where the constant $\kappa$ is a dimensionless parameter fixed to
2 for certain physical reasons. For the mass density signal, we have
the smeared form 
\begin{equation}
m_{\sigma,t}(\mathbf{x})=<\hat{m}_{\sigma}(\mathbf{x})>_{t}+\delta m_{t}(\mathbf{x}),
\end{equation}
where $\delta m_{t}(\mathbf{x})$ now satisfies
\begin{equation}
\mathrm{E}\left[\delta m_{t_{2}}(\mathbf{x})\delta m_{t_{1}}(\mathbf{y})\right]=-\frac{1}{4\pi\kappa G}\delta(t_{2}-t_{1})\nabla^{2}\delta(\mathbf{x}-\mathbf{y}).
\end{equation}
To avoid obvious divergences, TD use the smeared density in $\hat{U}_{g,\sigma}$,
resulting in the SME:

\begin{equation}
\begin{split}\frac{d\hat{\rho}}{dt} & =-i\left[\hat{H}+\hat{U}_{g,\sigma}+\int d^{3}x\delta m(\mathbf{x})\hat{V}_{\sigma},\hat{\rho}\right]\\
 & -\frac{\kappa G}{8}\int\frac{d^{3}xd^{3}y}{|\mathbf{x}-\mathbf{y}|}\left[\hat{m}_{\sigma}(\mathbf{x}),\left[\hat{m}_{\sigma}(\mathbf{x}),\hat{\rho}\right]\right]\\
 & -\frac{1}{8\pi\kappa G}\int d^{3}x\left[\nabla\hat{V}_{\sigma}(\mathbf{x}),\left[\nabla\hat{V}_{\sigma}(\mathbf{x}),\hat{\rho}\right]\right]\\
 & +\frac{\kappa G}{2}\int\frac{d^{3}xd^{3}y}{|\mathbf{x}-\mathbf{y}|}\mathcal{H}\left[\hat{m}_{\sigma}(\mathbf{x})\right]\hat{\rho}\delta m(\mathbf{x}),
\end{split}
\end{equation}
where the gravitational back-action 
\begin{equation}
\hat{U}_{g,\sigma}=\int d^{3}x\hat{V}_{(\sigma)}(\mathbf{x})\hat{m}_{\sigma}(\mathbf{x})=-\frac{G}{2}\int d^{3}xd^{3}y\frac{\hat{m}_{\sigma}(\mathbf{x})\hat{m}_{(\sigma)}(\mathbf{y})}{|\mathbf{x}-\mathbf{y}|}.
\end{equation}
By combining the two decoherence terms in (66) and setting $\kappa=2$
on the requirement that decoherence be minimal, they obtain the local
SME: 

\begin{equation}
\begin{split}\frac{d\hat{\rho}}{dt} & =-i\left[\hat{H}+\hat{U}_{g,\sigma}+\int d^{3}x\delta m(\mathbf{x})\hat{V}_{\sigma},\hat{\rho}\right]\\
 & -\frac{1}{8\pi G}\int d^{3}x\left[\nabla\hat{V}_{\sigma}(\mathbf{x}),\left[\nabla\hat{V}_{\sigma}(\mathbf{x}),\hat{\rho}\right]\right]\\
 & -\int d^{3}x\mathcal{H}\left[\hat{V}_{\sigma}(\mathbf{x})\right]\hat{\rho}\delta m(\mathbf{x}).
\end{split}
\end{equation}
The back-action term thereby doubles the decoherence term present
in the original DP master equation (48). TD take their cut-off $\sigma=10^{-15}m$
to remedy the divergence problem, but the overheating problem remains.
Hence, it is necessary again to introduce dissipative terms as done
by Bahrami et al. While TD don't incorporate dissipative terms in
their equations, it is clear that doing so will lead us to Bahrami
et al.'s constraint (49), in turn leading us to the conclusion that
TD-DP should also be regarded as an effective theory valid only for
masses comparable to or greater than $\sim10^{11}amu$. 

So the dissipative generalization of the TD-DP theory should coincide
with the dissipative DPmN theory on the following prediction: for
a microsphere with $m\sim10^{14}amu$, a cat state of CM position
states formed by either parametric coupling to a qubit (Romero-Isart
et al. protocol) or by free expansion in a magnetic skatepark potential
(Pino et al. protocol) will be rapidly suppressed to a width of $10^{-15}m,$
on the timescale given by (52). So in the g-cat setup, the classical
probe will detect a force from a full-mass microsphere located in
one of the two potential well minima for virtually all times, as in
TD-CSL.

\subsubsection{K-model}

The collapse model of Karolyhazy (K-model) posits that intrinsic spacetime
fluctuations couple to quantum systems and induce a discrete-time
state-vector reduction process similar to the GRW process \cite{Karolyhazy1986,A.Bassi2013}.
In particular, the spacetime fluctuations are encoded in a family
of perturbed metrics $\left\{ g_{\mu\nu}^{\beta}\right\} $ very close
to the Minkowski metric; these metrics modify the \emph{N}-body Schrödinger
equation for free particles to the form 
\begin{equation}
i\hbar\partial_{t}\psi_{\beta}=\left(-\sum_{i=1}^{N}\frac{\hbar^{2}}{2m_{i}}\nabla_{i}^{2}+U_{\beta}\right)\psi_{\beta},
\end{equation}
where $U_{\beta}$ encodes small perturbations given by 
\begin{equation}
U_{\beta}(\mathbf{x},t)=\sum_{i}\frac{m_{i}c^{2}\gamma_{\beta}(\mathbf{x}_{i},t)}{2},
\end{equation}
and where $\gamma_{\beta}(\mathbf{x}_{i},t)$ encodes the spacetime
fluctuations that induce the state-vector reduction. 

Note that Eq. (72) has a straightforward SN analogue: we simply adopt
Eqs. (25-27) in subsection 3.1, make the replacement $\psi\rightarrow\psi_{\beta}$
, and add to the Hamiltonian of (27) the $U_{\beta}$ term. Then we
have an \emph{N}-body K-model with matter density ontology that includes
\emph{N}-body gravitational pair interactions along with \emph{N}
single-body gravitational self-interactions of SN type. Such an extended
K-model (which we will call the KmN-model) can then be used to describe
the g-cat setup. 

For a single elementary particle, the K-model gives the ``critical
width'' for a wavepacket as ($L_{p}\approx10^{-35}m$) 
\begin{equation}
a_{c}\approx\left(\frac{L}{L_{p}}\right)^{2}L,
\end{equation}
where $L\approx\hbar/mc$, and the ``critical time'' of reduction
\begin{equation}
\tau_{c}\approx\frac{ma_{c}^{2}}{\hbar}.
\end{equation}

If we compare the K-model to GRW, $a_{c}$ is analogous to $\sigma_{GRW}$
and $\tau_{c}$ is analogous to $\lambda_{GRW}^{-1}$. For a macroscopic
body with center of mass $m_{tot}=\sum_{i=1}^{N}m_{i}$ and size $R$,
it can be shown that Eq. (74) becomes
\begin{equation}
a_{c}\approx\left(\frac{R}{L_{p}}\right)^{2/3}L,
\end{equation}
where now $L\approx\hbar/m_{tot}c$. Hence, for a microsphere of $R\sim1\mu m$
and mass $m_{tot}\sim10^{14}amu\sim10^{-13}kg$, Eqs. (76) and (75)
give
\begin{equation}
a_{c}^{sphere}\sim10^{-11}m,
\end{equation}
and 
\begin{equation}
\tau_{c}^{sphere}\approx\frac{m_{tot}\left(a_{c}^{sphere}\right)^{2}}{\hbar}\sim0.1s.
\end{equation}
So the microsphere would undergo ten collapses in one second; and
like GRWmN and CSLmN, in the KmN-model, after the first collapse,
the SN self-interaction will prevent delocalization of the wavepacket,
making subsequent collapse events ineffectual to the subsequent width
of the wavepacket. 

The timescale (78) is out of the range of the Romero-Isart et al protocol
for preparing the coherent microsphere, and the critical width is
$a_{c}>10^{-12}m$, implying that a collapse event won't destroy the
cat state, entailing no net force on the probe in the g-cat setup.
On the other hand, the timescale (78) is well within the timescale
of Pino et al.'s protocol (which we recall has a coherence time on
the order of one second) \cite{H.Pino2016}. So we can predict that
if one collapse event happens before the microsphere interacts with
the double-slit barrier, the microsphere matter density can only then
pass through one of the slits (recall that the slit width $w=10.61nm$,
or three orders of magnitude larger than $a_{c}^{sphere}$) and end
up in only one of the minima of the double-well potential on the other
side of the barrier. Moreover, like in GRWmN and CSLmN, in the KmN-model
the SN self-interaction (and any subsequent collapse events) will
inhibit tunneling of the microsphere matter density between the two
minima. 

For a microsphere mass $m_{tot}\sim10^{18}amu$ (keeping $R$ still
$\sim1$ micron), we note that $\tau_{c}^{sphere}$ for a microsphere
prepared with the Pino et al.'s protocol would be the same as $\lambda_{GRW}$.
So the KmN-model's prediction for the g-cat setup is the same as that
of GRWmN.

\subsection{Collapse theories with quantized gravity}

Here we compare the predictions of the semiclassical gravitational
OCTs to the predictions we would obtain from treating the Newtonian
gravitational potential as a quantized field. To do this, we need
only analyze in detail the GRW case, since the results therein will
be readily applicable to the other OCTs.

To start off, we consider the GRW theory with no primitive ontology
(i.e., no matter density in space-time and no flashes), which we will
call ``GRW0'' \cite{S.Goldstein2012}. For GRW0-Newton, we will
treat the gravitational potential sourced by matter as an operator-valued
field satisfying the Poisson equation 
\begin{equation}
\nabla^{2}\hat{V}_{g}=4\pi G\sum_{i=1}^{N}m_{i}\delta^{(3)}(\mathbf{x}-\hat{\mathbf{R}}_{i}),
\end{equation}
where the right hand side is a sum over all the (first quantized)
mass density operators associated to each particle. Because we are
treating the gravitational potential as operator-valued, the Schrödinger
evolution in GRW0-Newton is linear in $\psi$, in contrast to the
SN evolution in GRWmN. Moreover, the classical gravitational potential
associated to $\hat{V}_{g}$ is obtained by taking the quantum expectation
value of both sides of (79):
\begin{equation}
\left\langle \Psi\right|\nabla^{2}\hat{V}_{g}\left|\Psi\right\rangle =\left\langle \Psi\right|\sum_{i=1}^{N}m_{i}\delta^{(3)}(\mathbf{x}-\hat{\mathbf{R}}_{i})\left|\Psi\right\rangle =4\pi G\sum_{i=1}^{N}\int d\mathbf{r}{}_{1}...d\mathbf{r}{}_{N}|\Psi(\mathbf{r}{}_{1},...,\mathbf{r}{}_{N},t)|^{2}m_{i}\delta^{(3)}(\mathbf{x}-\mathbf{r}{}_{i}),
\end{equation}
where $\Psi=\Psi(\mathbf{r}{}_{1},...,\mathbf{r}{}_{N},t)$ and the
position operators $\hat{\mathbf{R}}_{i}$ give $\hat{\mathbf{R}}_{i}\Psi=\mathbf{r}_{i}\Psi$.
Accordingly, for an N-body system of identical particles with Newtonian
gravitational interactions, the N-body Schrödinger equation of GRW0-Newton
is given by 
\begin{equation}
i\hbar\partial_{t}\Psi(\mathbf{r}{}_{1},...,\mathbf{r}{}_{N},t)=\left[-\sum_{i=1}^{N}\frac{\hbar^{2}}{2m}\nabla_{i}^{2}-\sum_{i\neq j}\sum_{j}\frac{Gm^{2}}{|\hat{\mathbf{r}}_{i}-\hat{\mathbf{r}}_{j}|}\right]\Psi(\mathbf{r}{}_{1},...,\mathbf{r}{}_{N},t).
\end{equation}
And, of course, we have the GRW process wherein the solution of (81)
undergoes intermittent, discontinuous collapses of the form

\begin{equation}
\Psi_{T+}(\mathbf{r}_{1},...,\mathbf{r}_{N})=\frac{1}{C}g(\mathbf{r}_{i}-\mathbf{X})^{1/2}\Psi_{T-}(\mathbf{r}_{1},...,\mathbf{r}_{N}),
\end{equation}
with collapse width $\sigma_{GRW}$, collapse rate $N\lambda_{GRW}$,
and probability density

\begin{equation}
\rho(\mathbf{X})=C(\mathbf{X})^{2}=\int d\mathbf{r}'_{1}...d\mathbf{r}'_{N}g(\mathbf{r}'_{i}-\mathbf{X})|\Psi_{T-}(\mathbf{r}'_{1},...,\mathbf{r}'_{N})|^{2},
\end{equation}
where $i$ is chosen randomly from $1,...,N.$

It might be noticed that (81) is also the Schrödinger equation of
CQT-Newton. Thus, if we assume that the particles are weakly interacting,
then by imposing $\left|\Psi(t)\right\rangle =\underset{N\rightarrow\infty}{lim}e^{-\left(\frac{i}{\hbar}\right)\hat{H}_{quant}t}\otimes_{i=1}^{N}\left|\chi\right\rangle =\otimes_{i=1}^{N}\left|\chi(t)\right\rangle $,
we recover the mean-field equations
\begin{equation}
\nabla^{2}V_{g}=4\pi Gm|\chi(\mathbf{r},t)|^{2},
\end{equation}
and 
\begin{equation}
i\hbar\frac{\partial\chi(\mathbf{r},t)}{\partial t}=\left[-\frac{\hbar^{2}}{2m}\nabla^{2}-G\int d\mathbf{r}'\frac{m^{2}|\chi(\mathbf{r}',t)|^{2}}{|\mathbf{r}-\mathbf{r}'|}\right]\chi(\mathbf{r},t),
\end{equation}
for the collective variable $\chi(\mathbf{r},t)$. Notice that since
this approximation assumes that the many-body wavefunction can be
factorized as $\left|\Psi\right\rangle =\otimes_{i=1}^{N}\left|\chi\right\rangle $,
the collapse rate for $\chi(\mathbf{r},t)$ is just $\lambda_{GRW}$.
So the mean-field description of GRW0-Newton is effectively indistinguishable
from the mean-field description of CQT-Newton. And, like CQT-Newton,
the mean-field description leading to (84-85) is inadequate for modeling
our g-cat setup (because the factorization ansatz is a poor approximation
for microspheres). 

Instead, we must consider the microsphere CM Schrödinger equation
given by (14): 
\begin{equation}
i\hbar\frac{\partial\Psi(\mathbf{r}_{cm},t)}{\partial t}=\hat{H}_{cm}\Psi(\mathbf{r}_{cm},t)=\left[-\frac{\hbar^{2}}{2m_{tot}}\nabla_{cm}^{2}-\frac{GMm_{tot}}{\hat{D}}+U_{well}\right]\Psi(\mathbf{r}_{cm},t),
\end{equation}
where $m_{tot}=Nm$, $M$ is the probe mass, $\hat{D}=\sqrt{y_{cm}^{2}+|\mathbf{\hat{x}}_{cm}|^{2}}$,
$\mathbf{\hat{x}}_{cm}$ has eigenvalues $-\mathbf{L}/2$ and $+\mathbf{L}/2$,
and $\mathbf{y}_{cm}$ is the fixed (c-number) y-displacement of the
CM of the test mass from the CM of the probe. Then we can follow AH
in introducing the initial cat state 
\begin{equation}
|\Psi>=c_{+}|+>+c_{-}|->,
\end{equation}
which for a microsphere mass of $\sim10^{14}amu$ will have a collapse
rate of $10^{-2}s^{-1}$, or $10^{2}s^{-1}$ for a microsphere mass
of $\sim10^{18}amu$. As before, for the microsphere cat state produced
by the Romero-Isart et al. protocol, the collapses will be ineffectual
since $\mathbf{L}=1pm$ $\ll$ $\sigma_{GRW}=0.1\mu m$. But for the
microsphere cat state produced by the Pino et al. protocol, the collapses
will appreciably change the width of the cat state since now $\mathbf{L}=0.5\mu m$.

Recall how AH \cite{Anastopoulos2015} assumed that the Newtonian
gravitational interaction between the classical probe and the microsphere
acts as a projective measurement according to the usual quantum measurement
postulates; but because the usual quantum measurement postulates are
based on ambiguous notions like ``information extraction'' and ``macroscopic'',
this assumption was difficult to rigorously justify. For GRW0-Newton,
the Newtonian gravitational interaction between the classical probe
and the microsphere will indeed act as a projective measurement in
the sense that the probe-sphere gravitational coupling will drive
up the collapse rate of the microsphere cat state, and thus lead to
predictions in agreement with CQT-Newton as described by AH (apart
from minute differences in statistics due to the GRW process). 

To show this we must, however, describe the probe within the context
of GRW0-Newton as well \footnote{We are grateful to Dennis Dieks for suggesting the general outlines
of the ensuing argument}. In particular, we must attribute to the probe a CM wavefunction
in the `ready state' $\Phi_{0}(\mathbf{r}_{cm})$. Projecting the
cat state (87) onto the CM coordinate space gives $\Psi(\mathbf{r}_{cm})=c_{+}\Psi_{+}(\mathbf{r}_{cm})+c_{-}\Psi_{-}(\mathbf{r}_{cm})$,
indicating that $\Psi(\mathbf{r}_{cm})$ is not an eigenstate of the
CM position operator $\hat{\mathbf{R}}_{cm}$. Then the interaction
Hamiltonian in (86) implies that $\hat{U}_{probe-sphere}\left(\Psi_{+}\otimes\Phi_{0}\right)=\Psi_{+}\otimes\Phi_{+}$
and $\hat{U}_{probe-sphere}\left(\Psi_{-}\otimes\Phi_{0}\right)=\Psi_{+}\otimes\Phi_{-}$,
where $\Phi_{+}$ denotes the probe wavefunction `deflected' (correlated)
towards the position of the $+$ minimum and $\Phi_{-}$ denotes the
probe wavefunction deflected towards the position of the $-$ minimum.
By the linearity of (86), we then have the entangled state
\begin{equation}
\hat{U}_{probe-sphere}\left(\Psi\otimes\Phi_{0}\right)=c_{+}\Psi_{+}\otimes\Phi_{+}+c_{-}\Psi_{-}\otimes\Phi_{-},
\end{equation}
since the probe/pointer states are orthogonal, i.e., $\Phi_{+}\cdot\Phi_{-}\approx0$
(this follows from the assumption that the probe is sensitive enough
to the gravitational force from the cat state that its two possible
CM positional deflections are macroscopically distinct, i.e. separated
by a distance greater than $10^{-7}m$). Moreover, because (88) is
an entangled state, if one of the probe particles undergoes a GRW
hit described by (82), then the entire state (88) will collapse as
well. Thus the collapse rate of the probe-sphere CM wavefunction will
be $\left(N_{probe}+N_{sphere}\right)\lambda_{GRW}$, and the probability
density for collapse into either $\Psi_{+}\otimes\Phi_{+}$ or $\Psi_{-}\otimes\Phi_{-}$
will be given by 

\begin{equation}
C(\mathbf{X})^{2}=\int d^{3}\mathbf{r}_{cm}g(\mathbf{r}_{cm}-\mathbf{X})|c_{+}\Psi_{+}\otimes\Phi_{+}+c_{-}\Psi_{-}\otimes\Phi_{-}|^{2},
\end{equation}
where $\mathbf{X}=\{+\frac{\mathbf{L}}{2},-\frac{\mathbf{L}}{2}\}$.
We stress that these conclusions will apply to the g-cat setup using
either the Romero-Isart et al. protocol or the Pino et al. protocol,
in agreement with CQT-Newton.

Since the probe is assumed to be a macroscopic device composed of
a much larger number of particles than the microsphere (e.g., $N_{probe}\sim10^{23}$),
we should expect the collapse of (88) to be frequent enough that macroscopic
observers (such as experimentalists, who themselves will also correspond
to many-body wavefunctions evolving by the GRW0-Newton laws of motion,
and entangled with the probe) will `perceive' (through the macroscopic
deflections of the probe) the microsphere as undergoing seemingly
instantaneous quantum jumps between the $+$ and $-$ minima. Let
us estimate the rate of collapse, based on the assumption that the
probe is described by the Sankey et al. trampoline resonator \cite{Sankey2015}.
The actual trampoline (i.e., the part of the probe that plays the
role of the pointer) has a mass of only $4.0ng\sim10^{15}amu$, but
it is tethered to a much larger and more massive (square shaped) silicon
wafer. The wafer has thickness $675\mu m$ and width $3mm$, and solid
silicon has density $\rho_{Si}=2.33\frac{g}{cm^{3}}$. From these
values we can calculate that the wafer is composed of $\sim10^{20}$
nucleons. Since the wavefunction of the trampoline resonator and the
wavefunction of the wafer are strongly entangled, their joint many-body
wavefunction $\Phi_{0}$ (however complicated it looks) therefore
has a collapse rate of $\sim10^{20}\cdot\lambda_{GRW}=10^{4}s^{-1}$,
or around 10,000 collapses per second. (We neglect further increases
in the collapse rate due to entanglement with the thermal environment
since we assume that the resonator operates at a temperature of $14mK$
or lower, where its force sensitivity is at peak value.) So when the
wafer+resonator system (i.e., the probe) gets entangled with the microsphere
through gravitational coupling, this collapse rate will also apply
to (88) (the microsphere adds only $\sim10^{14}$ nucleons, which
negligibly increases the collapse rate).

Using the above analysis, we can also show without ambiguity why the
sphere's g-coupling to the Earth's field (and the field of any other
massive body in the environment) doesn't collapse the sphere's cat
state wavefunction, despite the magnitude of the Earth-sphere g-coupling
being nearly seventeen orders of magnitude greater than the probe-sphere
g-coupling. Suppose we take the CM of the Earth as the `pointer variable'
which correlates to the sphere states. We represent the CM of the
Earth by the ready state $\Phi_{0}^{Earth}(\mathbf{r}_{cm})$, and
replace the probe-sphere interaction term in (86) with the Earth-sphere
interaction term $-m_{tot}g\hat{z}$, where $\hat{z}$ is the operator-valued
vertical displacement of the CM of the sphere from the ground-level
of the lab. Then, following through the same argument leading to (88),
we have 
\begin{equation}
\hat{U}_{Earth-sphere}\left(\Psi\otimes\Phi_{0}^{Earth}\right)=c_{+}\Psi_{+}\otimes\Phi_{+}^{Earth}+c_{-}\Psi_{-}\otimes\Phi_{-}^{Earth}.
\end{equation}
This time, however, because of the huge mass of the Earth relative
to the sphere, the CM states $\Phi_{+}^{Earth}$ and $\Phi_{-}^{Earth}$
have considerable overlap and therefore are not orthogonal. Indeed,
the huge mass of the Earth implies that the relative separation between
$\Phi_{+}^{Earth}$ and $\Phi_{-}^{Earth}$ will be much less than
$\sigma_{GRW}$. That means $\Phi_{+}^{Earth}\approx\Phi_{-}^{Earth}\approx\Phi_{0}^{Earth}$,
and we can well approximate (90) as 
\begin{equation}
\hat{U}_{Earth-sphere}\left(\Psi\otimes\Phi_{0}^{Earth}\right)\approx\left(c_{+}\Psi_{+}+c_{-}\Psi_{-}\right)\otimes\Phi_{0}^{Earth}.
\end{equation}
Accordingly, when one of the Earth particles undergoes a GRW hit,
no change is entailed for the sphere's cat state. So the only particles
that are physically relevant to the collapse of the cat state are
the particles composing the sphere, and even then only if the relative
separation of the cat state components is greater than $\sigma_{GRW}$
(e.g., as in the Pino et al. protocol).

An objection that might be raised towards GRW0-Newton is that it is
an empirically incoherent theory because it predicts no space-time
and no matter in space-time to which experiments, observers, and the
perceptions of observers correspond (in stark contrast to our very
definite perceptions of living in a 3D-space with matter evolving
in it) \cite{Allori2012}. Indeed, the fundamental ontology of GRW0-Newton
is just an N-body wavefunction on configuration space. So when we
say that ``the probe wavefunction deflected towards the position
of the $-$ minimum'', what we really mean is that the probe-sphere
wavefunction collapses (effectively) to the sub-space of the probe-sphere
Hilbert space corresponding to the state $\Psi_{-}\otimes\Phi_{-}$.
Two possible answers to this objection are as follows: (i) Albert's
(philosophical) functionalist analysis of the GRW wavefunction \cite{Albert2015}
can be employed to deduce 3D-space and a matter density (or flash)
ontology within 3D-space as emergent ontological variables; (ii) we
can still postulate, in addition to an ontic wavefunction on configuration
space, the existence of a 3D-space and a matter density (or flash)
ontology within it, but with the understanding that these primitive
ontological variables are causally inert in space-time (i.e., the
matter density fields in space-time don't physically interact with
each other through classical forces, but only indirectly through the
evolution of the wavefunction in configuration space, and the flash
events don't get accompanied by point masses at the flash locations).
It is debatable which of these two options is more plausible than
the other (or, for that matter, if either option is plausible in its
own right), but for the purposes of this paper, we simply note that
they are both logically possible solutions to the `empirical incoherence'
objection.

By analogy with GRW0-Newton, it is straightforward to construct CSL0-Newton,
DP0-Newton, and K0-Newton. (Note that there is no TD0-Newton, since
the TD theory is specifically designed to treat the gravitational
field as classically sourced by the flash ontology in the setting
of CSL dynamics). Apart from minute differences in experiment statistics
resulting from the different intrinsic collapse rates predicted by
the CSL, DP, and K-model processes, it is straightforward to show,
using the same arguments as above, that these three variants of GRW0-Newton
will predict the same outcomes as GRW0-Newton for the g-cat setup.

\subsection{Related collapse theories}

While we have left out certain objective collapse theories from our
analyses above \cite{Adler2002,Adler2013,Kafri2014,Nimmrichter2015,Bera2015},
our findings up to this point allow us to quickly assess these other
ones.

The stochastic extension of the SN equations given by Nimmrichter
and Hornberger (NH) \cite{Nimmrichter2015} results in cancellation
of any gravitational self-interaction or pair interaction. Thus the
NH theory predicts \emph{no} gravitational coupling between probe
and microsphere, regardless of whether or not the microsphere can
be put into a coherent and stable spatial superposition. 

The theory of Kafri et al. \cite{Kafri2014} is mathematically and
conceptually equivalent to the original DP theory, which makes our
analysis of the DP theory applicable to their theory as well.

The theory of Bera et al. \cite{Bera2015} is formally equivalent
to the K-model in that it predicts the critical widths (74) and (76),
as well as the collapse timescale (75) and (78). (Although these timescales
are associated with a gravitationally-induced decoherence process
rather than a state-vector reduction process.) Thus our predictions
for the K-model apply as well to Bera et al.'s theory (modulo the
conceptual difference between a decoherence process and a state-vector
reduction process). 

Finally, the Trace Dynamics theory of Stephen Adler \cite{Adler2002}
results in an effective stochastic master equation that can (under
certain assumptions) be put into a form equivalent to that of the
CSL master equation. As such, our conclusions about CSL with semiclassical
gravitational pair interactions, whether in the form of CSLmN or TD-CSL,
will presumably also apply to Adler's semiclassical gravitational
generalization of Trace Dynamics \cite{Adler2013} (when one considers
the Newtonian limit).

\section{Conclusion}

We have appraised the most well-known and well-developed objective
collapse theories in light of DAH's proposed g-cat setup \cite{DerAnasHu2016},
including an extension of the g-cat setup to incorporate Pino et al.'s
protocol \cite{H.Pino2016}, and compared the predictions of said
collapse theories to the predictions of CQT-Newton. In particular,
we assessed the predictions of GRW, CSL, DP, and the K-model, in the
context of two cases: (i) extended to include semiclassical gravitational
interactions in the approach of Derakhshani in \cite{Derakhshani2014}
and/or the approach of Tilloy-Diósi in \cite{Tilloy2016}; and (ii)
extended to include quantized Newtonian gravitational interactions
between particles. We then used these results to assess other (closely
related) objective collapse theories in the recent literature, namely,
the theories of Nimmrichter \& Hornberger \cite{Nimmrichter2015},
Kafri et al. \cite{Kafri2014}, Bera et al. \cite{Bera2015}, and
Adler's Trace Dynamics \cite{Adler2002}. 

The results of our primary analyses can be summarized as follows:
\begin{enumerate}
\item \textbf{GRWmN:} (i) The probe-microsphere (or even Earth-microsphere)
semiclassical gravitational coupling for the g-cat setup will not
drive up the collapse rate of the microsphere; (ii) the number of
nucleons composing the microsphere is too few to bring its collapse
rate within the coherence time of the g-cat setup using the Romero-Isart
et al. protocol, and in any case the relative separation of $\mathbf{L}=1pm$
is much smaller than $\sigma_{GRW}$, thereby implying that any collapse
event will make no physical change to the sphere's cat state wavefunction
and associated matter density (so the probe will just feel net-zero
gravitational force from an uncollapsed cat state matter density);
(iii) for the g-cat setup using the Pino et al. protocol (which makes
possible $\mathbf{L}\gtrsim0.5\mu m$), the microsphere mass can potentially
be increased by as much as four orders of magnitude, thereby bringing
the sphere's collapse rate well within the coherence time of the protocol
and making it possible that the force probe could measure a GRW-type
quantum jump of the gravitational force from the microsphere cat state,
or else just a continuous gravitational force from the microsphere
being localized to one of the minima of the double-well potential
illustrated in Fig. 1 (with no probability of tunneling between the
minima, due to SN self-interaction). 
\item \textbf{CSLmN: }(i) The probe-microsphere g-coupling for the g-cat
setup will not drive up the collapse rate of the microsphere; (ii)
however, the rate of effective localization on the center-of-mass
motion of the microsphere will be so high that a microsphere cat state
prepared using Pino et al.'s protocol will be quickly suppressed and
remain suppressed thereafter (due to SN self-interaction), while for
a microsphere cat state prepared with the Romero-Isart et al. protocol
the cat state will remain stable for the coherence time of the experiment;
(iii) thus, if the g-cat setup were experimentally implemented using
the Pino et al. protocol, the probe would measure a continuous gravitational
force from a full-mass microsphere matter density that sits in one
of the minima of a double-well potential for all times, while use
of the Romero-Isart et al. protocol would entail net-zero gravitational
force on the probe from an uncollapsed cat state matter density.
\item \textbf{DPmN: }Corrected with a cut-off and the inclusion of dissipative
terms to prevent overheating, DPmN predicts: (i) that the probe-microsphere
g-coupling for the g-cat setup will not drive up the collapse rate
of the microsphere; (ii) for the microsphere, a characteristic damping
time (i.e., state-vector reduction rate) even more rapid than that
of CSL (two orders of magnitude more, to be exact); (iii) rapid collapse
of the microsphere cat state in both the Romero-Isart et al. protocol
(because the spatial cut-off $R_{0}=10^{-15}m$ vs. $\mathbf{L}=1pm$)
and the Pino et al. protocol; (iv) thus, for both the Romero-Isart
et al. protocol and the Pino et al. protocol, the probe in the g-cat
setup would measure a continuous gravitational force from a full-mass
microsphere matter density that sits in one of the minima of the double-well
potential for all times (due to SN self-interaction). 
\item \textbf{TD-CSL/DP: }The CSL case predicts: (i) semiclassical g-coupling
to the probe or any other massive body in the environment doesn't
drive up the collapse rate of the microsphere; (i) extremely slow
phase damping (decoherence) of the microsphere density matrix due
to gravitational back-action; (ii) the same collapse rate and collapse
width for the microsphere as in CSLmN (and ordinary CSL), which means
collapse events make a physical difference for the microsphere prepared
by the Pino et al. protocol and no physical difference for the Romero-Isart
et al. protocol. But because there is no SN self-interaction, there
is a small probability of the cat state tunneling between the minima
in between collapse events. However, this tunneling rate is so low
that a tunneling event is unlikely to be observed on the timescales
of the g-cat experiment. So TD-CSL makes effectively the same prediction
for the g-cat setup as CSLmN and DPmN, for both the Pino et al. protocol
and the Romero-Isart et al. protocol. Similarly, the DP case predicts:
(i) same as TD-CSL; (ii) doubling of the decoherence term in the original
DP master equation; and (iii) when corrected with the appropriate
length-scale cut-off and dissipative terms, the same characteristic
damping time as the original dissipative DP(mN) theory. And like TD-CSL,
the tunneling rate in between collapse events is negligible for the
g-cat experiment. So TD-DP makes effectively the same predictions
for the g-cat setup as DPmN.
\item \textbf{KmN-model: }(i) semiclassical g-coupling to the probe or any
other massive body in the environment doesn't drive up the collapse
rate of the microsphere;\textbf{ }(ii) the predicted microsphere collapse
rate is on the order of a tenth of a second (for $m\sim10^{14}amu$),
which is out of the range of the coherence time of the microsphere
for the Romero-Isart et al. protocol, and in any case a collapse event
will yield no physical change to the sphere's cat state wavefunction
and associated matter density since $a_{c}^{sphere}\sim10^{-11}m$,
so no net force on the probe in the corresponding g-cat setup; (iii)
however, the collapse rate falls within the coherence timescale of
the Pino et al. protocol, and for that rate makes essentially the
same predictions as GRWmN for the g-cat setup. 
\item \textbf{GRW0-Newton:} (i) The probe-microsphere gravitational coupling
for the g-cat setup \emph{will} drive up the collapse rate of the
microsphere, and thereby result in the same predictions as CQT-Newton
for the case of a classical probe continuously monitoring the g-field
of the microsphere cat state, for both microsphere-preparation protocols
(apart from minute differences in the g-cat experiment statistics
entailed by the GRW process); (ii) but the g-coupling of the sphere
to other massive bodies in the environment, such as the Earth, \emph{will
not} drive up the collapse rate of the sphere, unless those other
massive bodies satisfy the physical conditions required to play the
role of a pointer variable (as in the case of the probe); and (iii)
apart from small differences entailed by the different intrinsic collapse
rates predicted by the CSL, DP, and K-model processes, CSL0-Newton,
DP0-Newton, and K0-Newton will predict the same outcomes as GRW0-Newton
for the g-cat setup. 
\end{enumerate}
We therefore conclude that the g-cat setup is, in principle, capable
of: (i) experimentally discriminating between the predictions of the
aforementioned semiclassical gravitational OCTs, to the extent that
some of these OCTs make predictions that differ from each other for
the g-cat setup; and (ii) experimentally discriminating between some
or all of the predictions of the aforementioned semiclassical gravitational
OCTs versus the predictions of CQT-Newton and GRW0/CSL0-DP0-K0-Newton,
to the extent that the analyzed semiclassical gravitational OCTs make
different predictions by virtue of treating the gravitational field
semiclassically instead of (perturbatively) quantized. 

For the purpose of solidifying the theoretical foundations of the
OCT theories analyzed here, it seems prudent to extend the semiclassical
gravitational OCTs to the regime of semiclassical Einstein gravity
(if possible!), take the Newtonian limit, and compare the resulting
predictions for the g-cat setup to the predictions obtained in this
paper. Likewise, to extend GRW0, CSL0, DP0, and K0-model to the regime
of relativistic perturbatively quantized gravity, take the non-relativistic
limit, and compare the predictions for the g-cat setup to the predictions
obtained in this paper. These are tasks for future work.

\section{Acknowledgments}

I am indebted to Bei-Lok Hu, Guido Bacciagaluppi, Dennis Dieks, and
Charis Anastopoulos for helpful discussions and useful feedback on
the topics in this paper. 

\bibliographystyle{unsrt}
\bibliography{/Users/maaneliderakhshani/RECOVERED/Users/maaneliderakhshani/Documents/GravCatProjectNotes,GRWm_refs}

\end{document}